\documentclass[a4paper,fleqn,usenatbib]{mnras}

\usepackage{newtxtext,newtxmath}

\usepackage[T1]{fontenc}
\usepackage{ae,aecompl}

\usepackage{graphicx}	
\usepackage{amsmath}	
\usepackage{amssymb}	

\usepackage{float}

\AtBeginShipout{%
  \ifnum\value{page}>1 %
    \typeout{* Additional boxing of page `\thepage'}%
    \setbox\AtBeginShipoutBox=\hbox{\copy\AtBeginShipoutBox}%
  \fi
}

\title[Internal systematics from input physics and surface correction methods]
{ Asteroseismic modelling of solar-type stars: internal systematics from input physics and surface correction methods } 
\author[B. Nsamba et al.]{
B. Nsamba,$^{1,2}$\thanks{E-mail: benard.nsamba@astro.up.pt}
T. L.  Campante, $^{1,2}$
M. J. P. F. G. Monteiro, $^{1,2}$
M. S. Cunha, $^{1,2}$
\newauthor B. M. Rendle, $^{3,4}$ D. R. Reese, $^5$ and K. Verma $^{4}$
\\
$^{1}$Instituto de Astrof\'{\i}sica e Ci\^{e}ncias do Espa\c{c}o, Universidade do Porto,  Rua das Estrelas, PT4150-762 Porto, Portugal\\
$^{2}$Departamento de F\'{\i}sica e Astronomia, Faculdade de Ci\^{e}ncias da Universidade do Porto, Rua do Campo Alegre, s/n, PT4169-007 Porto, \\Portugal\\
$^{3}$School of Physics and Astronomy, University of Birmingham, Edgbaston, Birmingham B15 2TT, UK\\
$^{4}$Stellar Astrophysics Centre, Department of Physics and Astronomy, Aarhus University, Ny Munkegade 120, DK-8000 Aarhus C, Denmark\\
$^{5}$LESIA, Observatoire de Paris, PSL Research University, CNRS, Sorbonne Universit\'{e}s, UPMC Univ. Paris 06, Univ. Paris Diderot,\\ Sorbonne Paris Cit\'{e}, 92195 Meudon, France
}

\date{Accepted 2018 April 12. Received 2018 April 12; in original form 2017 December 8}

\pubyear{2018}

\begin{document}
\label{firstpage}
\pagerange{\pageref{firstpage}--\pageref{lastpage}}
\maketitle

\begin{abstract}
Asteroseismic forward modelling techniques are being used to determine fundamental properties (e.g.  mass, radius, and age) of solar-type stars.  The need to take into account all possible sources of error is of paramount importance towards a robust determination of stellar  properties. We present a study of 34 solar-type 
stars for which high signal-to-noise asteroseismic data is available from multi-year \textit{Kepler} photometry. We explore the internal systematics on the stellar properties, that is, associated with the uncertainty in the input physics used to construct the stellar models. In particular, we explore the systematics arising from:
(i) the inclusion of the diffusion of helium and heavy elements; and (ii) the uncertainty in solar metallicity mixture.
We also assess the systematics arising from (iii) different surface correction methods used in optimisation/fitting procedures.
The systematics arising from comparing results of models with and without diffusion are found to be 0.5\%, 0.8\%, 2.1\%, and 16\% in mean density, radius, mass, and age, respectively. The internal systematics in age are significantly larger than the statistical uncertainties. 
We find the internal systematics resulting from the uncertainty in solar metallicity mixture to be 0.7\% in mean density, 0.5\% in radius, 1.4\% in mass, and 6.7\% in age.  The surface correction method by \citeauthor{Sonoi} and \citeauthor{Ball}'s two-term correction produce the lowest internal systematics among the different correction methods, namely, $\sim$1\%, $\sim$1\%, $\sim$2\%, and $\sim$8\% in mean density, radius, mass, and age, respectively. Stellar masses obtained using the surface correction methods by \citeauthor{Kjeld} and \citeauthor{Ball}'s one-term correction are systematically higher than those obtained using frequency ratios.
\end{abstract}




\section{Introduction}

Our knowledge of the underlying physical processes taking place in deep stellar interiors is of great importance for the accurate characterization of stars and classification of stellar populations. 
The treatment and choice of the essential model input physics such as solar metallicity mixture \citep{Grevesse,Asplund,Lodders}, initial helium abundance \citep{Chiosi,Casagrande}, as well as the different mixing processes like convection, semi-convection, convective overshooting \citep{Monteiro,Thomps,Deheuvels,Piau,Silva,Trampedach}, microscopic diffusion, radiation levitation, and rotational mixing \citep{Thoul, Turcotte, Maeder} have a direct impact on the derived stellar parameters such as mass, radius, and age. 

In order to aid the characterization of extra-solar planetary systems, current efforts are being geared towards improving the synergies between planetary science and asteroseismology (see, e.g. the book by \citealt{SCM}).  Asteroseismology has proven to be an effective technique in determining precisely the exoplanet-host star parameters and hence planet properties \citep{Huber,Beno,Marcy,Camp,CTL}.
Transit observations can only provide an estimate of the planet-to-star radius ratio. Therefore, precise stellar radii from asteroseismology allow tight constraints to be placed on the absolute sizes of planets. For bright enough systems, radial-velocity observations may be combined with the transit data to estimate planetary masses.
The inferred planetary mass scales with
the stellar mass according to $M_{\rm p} \propto M^{2/3}$ \citep[e.g.][]{Perryman2014}
which asteroseismology can again provide \citep{SCM}.
Last but not least, stellar ages from asteroseismology can potentially be used to assess the dynamical stability of planetary systems and to establish their relative chronology.

Photometric observations by NASA's \textit{Kepler} space telescope \citep{Borucki} led to the characterization of several hundred solar-type stars using asteroseismology. Future space missions such as NASA's Transiting Exoplanet Survey Satellite \citep[TESS;][]{Campante} will allow the detection of oscillations in up to 10{,}000 solar-type stars with low temporal resolution, whereas ESA's PLAnetary Transits and Oscillations of stars mission \citep[PLATO;][]{Rauer1} is expected to reach 
$\sim$80{,}000 solar-type stars with detected oscillations based on multi-year observations. With this in mind, efforts are being directed towards increasing the precision of asteroseismic inferences by matching the observed oscillation frequencies (or their combinations) to the corresponding frequencies (or their combinations) obtained from the stellar evolutionary models  \citep{Monta,Still,Metcalfe,Kawaler,Aguirre,Dav}. This approach is known to improve the precision of derived stellar parameters over forward modelling methods that only consider global oscillation parameters \citep[i.e. the frequency of maximum oscillation power, $\nu_{\rm max}$, and the large frequency separation, $\Delta \nu$;][]{Mathur,Lebreton}. It nevertheless yields stellar parameters that are model-dependent and therefore sensitive to the input physics used in the models. For instance, the estimated stellar ages are sensitive to different transport processes such as microscopic diffusion, convection and overshooting, which need to be parameterized. Consequently, the treatment of the input physics becomes a source of uncertainty that cannot be easily accounted for.

\cite{Aguirre} compared stellar properties of 33 \textit{Kepler} planet-candidate host stars derived using a variety of stellar evolutionary codes and optimisation/fitting methods, yielding internal systematics of $\sim$1\% in radius and density, $\sim$2\% in mass, and $\sim$7\% in age. In order to avoid internal systematics arising from the adoption of a variety of evolution and optimisation tools, we employ the same tools in all computations performed in this work.
This paper is then aimed at exploring systematic effects arising from specific choices of the input physics used in models of solar-type stars. In particular, we explore internal systematics\footnote{Hereafter, we describe the internal systematics as the scatter, $\sigma$, induced on the derived stellar parameters from differences in the input physics or the surface correction method.} arising from the inclusion of diffusion in model grids and changes in element abundances. Inclusion of atomic diffusion in stellar models and its impact on the derived stellar parameters has been the subject of a number of studies over the past decades \citep{Aller,VandenBerg,Dotter}. This is mainly because atomic diffusion has been revealed to occur in the Sun and other stars \citep{Guzik,Christensen,Korn}. We also explore the internal systematics arising from the uncertainty in the solar metallicity mixture. Different solar metallicity mixtures \citep[e.g.][]{Grevesse, Asplund} are being adopted in stellar modelling tools despite differences in the absolute element abundances \citep{Monta,Serenelli,Aguirre,Aguirre1}. This hence becomes a potential source of uncertainty in the derived stellar parameters. 
Furthermore, we assess the internal systematics arising from commonly used surface correction methods. \citet{WHG} investigated the performance of different surface correction methods applied to evolved stars (i.e. subgiants and low-luminosity red giants) and established the total additional uncertainties in the derived radii, masses, and ages to be less than 1\%, 2\%, and 6\%, respectively. In this paper, we assess the performance of different surface correction methods by comparing them with the use of frequency ratios, known to be less prone to near-surface effects \citep{Rox}.

This paper is organised as follows. In Sect.~\ref{sample}, we describe our target sample, as well as the adopted seismic and spectroscopic observables. In Sect.~\ref{grid}, we present the stellar evolution code, the different model grids, and the optimisation procedure used, while the main results are discussed in Sect.~\ref{results}. Section \ref{con} presents a summary of our findings.

\section{Target Sample}
\label{sample}

Our sample consists of 34 solar-type oscillators which have been observed by the \textit{Kepler} satellite. Of these, 32 stars are part of the `LEGACY' sample \citep{Lund,Aguirre1} with the remaining 2 stars being the components of the asteroseismic binary HD~176465 \citep{white,Ben}. 
These stars were observed in \textit{Kepler} short-cadence mode ($\Delta t = 58.89\:{\rm s}$) for at least 12 months. The sample includes some of the highest signal-to-noise ratio, solar-like oscillators observed by \textit{Kepler}.
Details about light curve preparation, power spectrum calculation, the peak-bagging procedure and adopted individual oscillation frequencies are given in \cite{Lund}. 

\begin{figure}
	\includegraphics[width=\columnwidth]{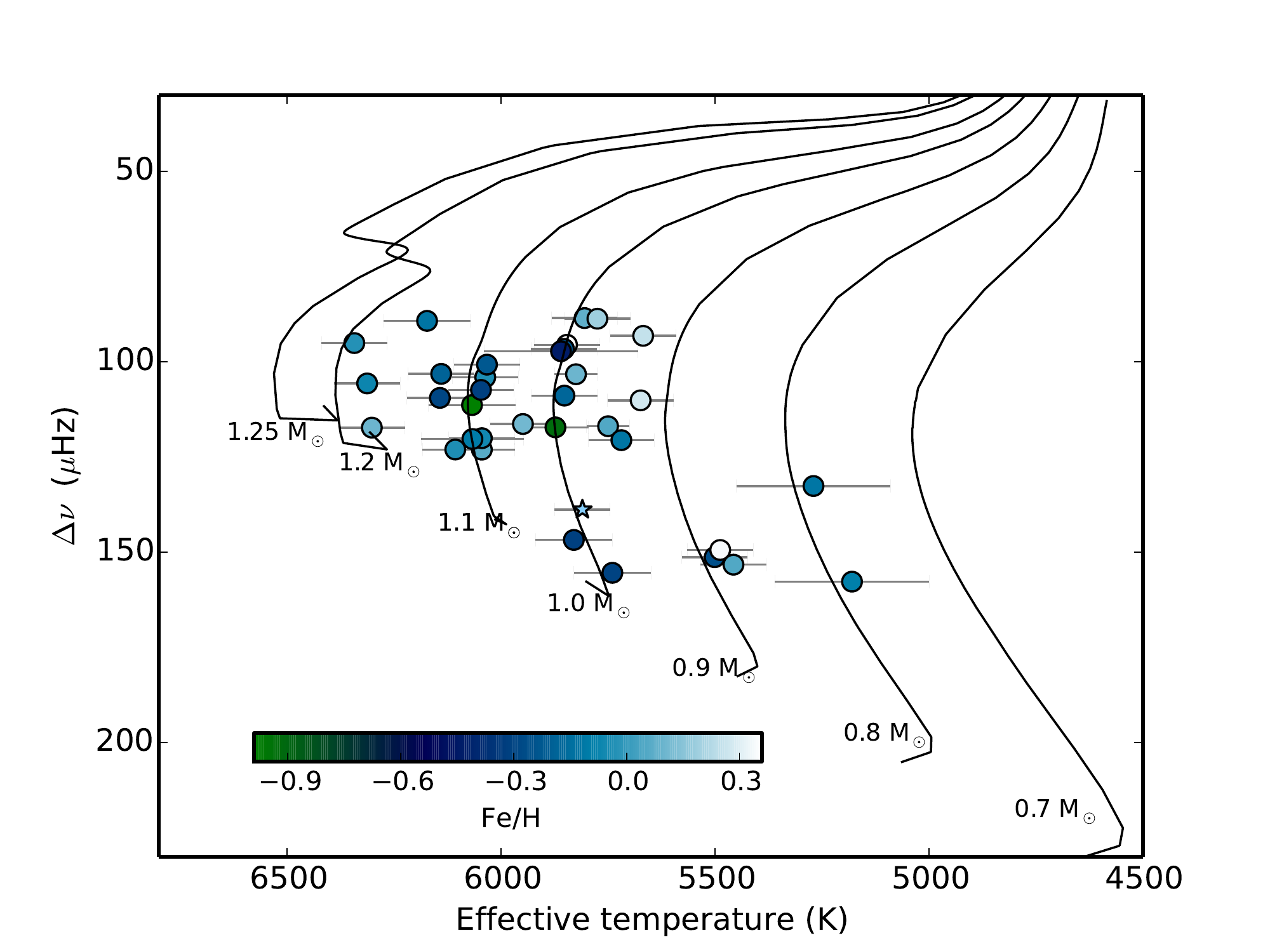}
      \caption{Target sample. Stellar evolutionary tracks were constructed at solar metallicity and range in mass from 0.7 to 1.25 M$_\odot$. Stars are
	colour-coded according to their metallicities. The `star' symbol corresponds to the position of the Sun.}
    \label{tracks}
\end{figure}

The target sample is shown in an asteroseismic Hertzsprung--Russell diagram in Fig.~\ref{tracks}. The adopted $\Delta\nu$ is from \cite{Lund}, computed following a Gaussian-weighted linear fit to $l = 0$ mode frequencies expressed as a function of the radial order. Most of the stars in the target sample are more evolved than the Sun. 
Table \ref{1} contains the atmospheric properties of the stars in our sample.
Most of the spectroscopic parameters were retrieved from \cite{Lund}. These classical constraints will complement the asteroseismic parameters in the optimisation procedure (as described in
Sect.~\ref{grid}). The components of the binary HD~176465 \citep{white} have effective temperatures $5830 \pm 90 \:{\rm K}$ (HD~176465 A) and $5740 \pm 90 \:{\rm K}$ (HD~176465 B), and similar metallicity ($-0.30 \pm 0.06 \:{\rm dex}$).
our
\begin{table}
\centering
\caption{Spectroscopic parameters of sample stars. The effective temperature ($\rm T_{\rm eff}$) and metallicity ([Fe/H]) are adopted from $^a$\citet{Casa}, $^b$\citet{Pinsonneault},$^c$\citet{Pinso}, and $^d$\citet{Ram} as indicated. The remaining parameters are from \citet{Lund}.}
\label{1}
\begin{tabular}{l c  r  }        
\hline 
KIC  &	$T_{\mbox{eff}}$ (K)	&  [Fe/H] (dex)	\\
\hline
3427720	&	6045 $\pm$ 77		&  -0.06 $\pm$ 0.10  	\\
3656476	&	5668 $\pm$ 77		&   0.25 $\pm$ 0.10 	\\
3735871	&	6107 $\pm$ 77		&  -0.04 $\pm$ 0.10 	 \\
4914923	&	5805 $\pm$ 77		&   0.08 $\pm$ 0.10 	\\
5184732	&	5846 $\pm$ 77		&   0.36 $\pm$ 0.10 	\\
5950854	&       5853 $\pm$ 77		&  -0.23 $\pm$ 0.10 	\\
6106415	&	6037 $\pm$ 77		&  -0.04 $\pm$ 0.10 	\\
6116048	&	6033 $\pm$ 77		&  -0.23 $\pm$ 0.10 	\\
6225718	&	6313 $\pm$ 77		&  -0.07 $\pm$ 0.10 	\\
6603624	&	5674 $\pm$ 77		&  0.28  $\pm$ 0.10 	\\
7106245	&	6068 $\pm$ 102$^a$	&  -0.99 $\pm$ 0.19$^a$  	\\
7296438	&	5775 $\pm$ 77		&  0.19  $\pm$ 0.10 	\\
7871531	&	5501 $\pm$ 77		&  -0.26 $\pm$ 0.10 	\\
8006161	&	5488 $\pm$ 77		&  0.34  $\pm$ 0.10 	\\
8150065	&	6173 $\pm$ 101$^a$	&  -0.13 $\pm$ 0.15$^a$ \\
8179536	&	6343 $\pm$ 77		&  -0.03 $\pm$ 0.10 \\
8379927	&	6067 $\pm$ 120$^b$ 	&  -0.10 $\pm$ 0.15$^b$ 	\\
8394589	&	6143 $\pm$ 77		&  -0.29 $\pm$ 0.10 	\\
8424992	&	5719 $\pm$ 77		&  -0.12 $\pm$ 0.10       \\
8760414	&	5873 $\pm$ 77		&  -0.92 $\pm$ 0.10 	\\
9025370	&	5270 $\pm$ 180$^c$	&  -0.12 $\pm$ 0.18$^c$ 	\\
9098294	&	5852 $\pm$ 77		&  -0.18 $\pm$ 0.10 	\\
9139151	&	6302 $\pm$ 77		&  0.10  $\pm$ 0.10 	\\
9410862	&	6047 $\pm$ 77		&  -0.31 $\pm$ 0.10 	\\
9955598	&	5457 $\pm$ 77		&  0.05  $\pm$ 0.10 	\\
9965715	&	5860 $\pm$ 180$^c$	&  -0.44 $\pm$ 0.18$^c$	\\
10079226 &	5949 $\pm$ 77		&  0.11 $\pm$ 0.10  	\\
10644253 &	6045 $\pm$ 77		&  0.06  $\pm$ 0.10 	\\
10963065 &	6140 $\pm$ 77		&  -0.19 $\pm$ 0.10 	\\
11772920 &	5180 $\pm$ 180$^c$	&  -0.09 $\pm$ 0.18$^c$ \\
12069424 &	5825 $\pm$ 50$^d$	&  0.10  $\pm$ 0.03$^d$ \\
12069449 & 	5750 $\pm$ 50$^d$  	&  0.05	 $\pm$ 0.02$^d$ \\
\hline                                   
\end{tabular}

\end{table}

\section{Stellar models and fitting procedure}
\label{grid}
We describe the stellar evolution code adopted in this work and the physics used in the construction of the different grids in Sect.~\ref{grid1}, while the adopted  optimisation/fitting procedure is described in Sect.~\ref{grid2}.

\subsection{Grid construction}
\label{grid1}
We used the 1D stellar evolution code Modules for Experiments in Stellar Astrophysics (MESA; \citealt{Pax1,Pax2,Pax3}) to generate grids of main-sequence and subgiant stellar models.
The evolutionary tracks were varied  in mass, $M$, initial metal mass fraction, $Z$, and mixing length parameter, $\alpha_{\rm mlt}$ \citep[][]{Vitense}. The parameter ranges are: $M \in$ [0.70, 1.25] M$_\odot$ in steps of 0.05 M$_\odot$, $Z \in$ [0.006, 0.031] in steps of 0.001, and $\alpha_{\rm mlt} \in$ [1.3, 2.9] in steps of 0.1. In our grids, neither convective overshoot nor semi-convection was included. Stellar models with masses $\gtrapprox$1.1 M$_\odot$ (at solar metallicity) are expected to have convective cores while on the main sequence and core overshoot may therefore be an important aspect to consider in the construction of a grid. Most of our target stars have masses below this mass limit. The impact of including  core overshoot is beyond the scope of this study and it has not been included in our grids. In two of the grids (see Table \ref{2}), element diffusion was included according to \cite{Thoul} to allow for diffusion of hydrogen and gravitational settling of heavy elements (i.e. $^4$He, $^{16}$O, and $^{56}$Fe). No radiative levitation was included in the models. 

Specifically, we used MESA version 7624, whose equation of state works with density, $\rho$, and temperature, $T$, as independent natural variables in a Helmholtz free energy formulation of thermodynamics. The basic input physics used in all of our grids includes the 2005 updated version of the OPAL equation of state \citep{Rogers}. The stellar models
used opacities from OPAL tables \citep{Iglesias} at high temperatures, whereas at lower temperatures tables from \cite{Ferguson} were used instead. Nuclear reaction rates were obtained from tables provided by the NACRE collaboration \citep{Angulo}. Specific rates for 
$^{14}{\rm N}(p,\gamma)^{15}{\rm O}$ were from \cite{Imbriani} and for $^{12}{\rm C}(\alpha, \gamma)^{16}{\rm O}$ from \cite{Kunz}. The standard Grey--Eddington atmosphere was used to integrate the atmospheric structure from the photosphere to an optical depth of $10^{-4}$. The initial helium mass fraction, $Y$, of our evolution models was determined using the helium-to-heavy
metal enrichment law anchored to the big bang nucleosynthesis values of $Z_{0}$ = 0.0 and $Y_{0}$ = 0.2484 \citep{Cyburt} . We therefore define the initial helium mass fraction according to 
\begin{equation}
 Y = \left(\frac{\Delta Y}{\Delta Z}\right)Z + Y_0 ~~.
	\label{mass_fraction}
\end{equation}
The enrichment law ratio, $\Delta Y  / \Delta Z$, ranges from 1 to 3 based on both theoretical and observational studies \citep{Jimenez,Balser,Casagrande,Serenelli}. Typically, a value of $\Delta Y  / \Delta Z$ determined through a solar calibration is adopted in Eq.~(\ref{mass_fraction}). This would mean that stars in the sample are assumed to have formed in regions having the same helium-to-heavy element mass fraction as the Sun. In order to avoid any systematics that could arise from variations in the treatment of the initial helium mass fraction, we set $\Delta Y  / \Delta Z$ = 2 \citep{Chiosi,Casagrande}  in all of our grids.
The systematic contributions arising from the treatment of initial helium mass fraction will be addressed in a separate paper.
\begin{table*}
\centering 
\caption{Summary of adopted grids.} 
\begin{tabular}{lccccr}        
\hline 
Name &	Mass Range (M$_\odot$)	&  Solar metallicity mixture	&  $\frac{\Delta Y}{\Delta Z}$ & Overshoot & Diffusion\\
\hline
GS98{\small sta} & 0.70 -- 1.25	& \cite{Grevesse} &	2.0			&	No	&  Yes	    \\
GS98{\small nod} & 0.70 -- 1.25	& \cite{Grevesse} &	2.0			&	No	&  No	    \\
AGS09		    & 0.70 -- 1.25	& \cite{Asplund}  &	2.0			&	No	&  Yes	     \\	
\hline                                   
\end{tabular}
\label{2}
\end{table*}

The grids were evolved starting from the pre-main sequence (PMS) to the zero-age main-sequence (ZAMS). We define the ZAMS as the point along the evolutionary track where the nuclear 
luminosity of the model yields 90\% of the total luminosity. All PMS models were discarded since our target stars are more evolved.
We then evolved the models from the ZAMS to the point along the evolutionary track where $\log\rho_{\rm c} = 4.5$ ($\rho_{\rm c}$ is the central density). This approximately corresponds to the base of the red-giant branch. About 70 models were stored at different ages along each evolutionary track and a total of about 371,280 models for each grid.
For each model, we used GYRE \citep{Townsend} in its adiabatic setting to generate theoretical oscillation frequencies. Pressure-mode (p-mode) oscillation frequencies were computed for harmonic degrees $l$ = 0, 1, 2, and 3 below the acoustic cut-off frequency.

It is worth noting that an offset is always seen between model and observed frequencies \citep{Dalsgaard1,Dziembowski,Dalsgaard}. This is due to an improper modelling of the near-surface layers. In order to model convection, the mixing-length theory is often used, which is only valid in the deep stellar interior and does not properly describe convection near the surface. In addition, the description of the interaction between oscillations and convection is still poorly understood. Also, other active processes like magnetic fields affect the properties of the oscillations and the equilibrium structure, however their inclusion in the modelling is challenging and thus usually neglected. Altogether, these give rise to a surface effect not properly accounted for in models, hence becoming a substantial obstacle to the direct comparison of model frequencies with observed frequencies. The surface effect in our model frequencies was corrected using various surface correction methods (see Sect.~\ref{freq}) and implemented in the optimisation tool (as described in Sect.~\ref{grid2}).

\subsection{Optimisation procedure}
\label{grid2}
AIMS\footnote{http://bison.ph.bham.ac.uk/spaceinn/aims/} (Asteroseismic Inference on a Massive Scale; Rendle et al. in prep) is based on a Bayesian approach and generates probability distribution functions (PDFs) of the different stellar parameters. In order to generate a representative set of 
models reproducing a specified set of asteroseismic and classical constraints, AIMS uses a Markov chain Monte Carlo (MCMC) algorithm based on \cite{Foreman} in combination with interpolation based on a Delaunay tessellation of the stellar grid.
With $A$ representing different stellar parameters and $O$ various asteroseismic and classical observables, from Bayes's theorem one has:
\begin{equation}
  p(A|O) \propto p(O|A) \, p(A) ~,
 \label{bayes}
\end{equation}
where $p(A)$ denotes our prior assumptions. We assigned uniform prior distributions to $M$, $Z$, and $\alpha_{\rm mlt}$. The likelihood of obtaining a set of observables given a set of model parameters is given by (see, e.g. the book by \citealt{Gregory})
\begin{equation}
  p(O|A) = \frac{1}{(2\pi)^\frac{1}{2} \sqrt{\lvert C \rvert}} \exp(-\chi^2/2) ~,
  \label{likelihood}
\end{equation}
where $C$ is the covariance matrix of the observed parameters.
It is worth noting that we assumed Gaussian distributed errors for our observables (i.e. individual oscillation frequencies, effective temperature, and metallicity). When dealing with independent observables, then $\chi^2$ is defined as 
\begin{equation}
 \chi^2 = \sum^N _{i=1} \left(\frac{O_i - \theta_i}{\sigma_i}\right)^2 ~,
\end{equation}
where $O_i$, $\theta_i$, and $\sigma_i$ are the observed value, modelled value, and the associated observed uncertainties respectively. When one instead fits frequency ratios (Sect.~\ref{freq}), correlations will be introduced that are a function of frequency. This is taken into account in the likelihood function (Eq.~\ref{likelihood}) and, in this case, the $\chi^2$ of each model is given by (e.g. \citealt{Gregory})
\begin{equation}
 \chi^2 = (O - \theta)^T C^{-1}(O - \theta)~.
\end{equation}
We stress that the frequency ratios used in this paper were calculated using AIMS. We further complemented the frequency ratios with the large frequency separation calculated from $l = 0$ modes as seismic constraints. Furthermore, we note that we give equal weights to the seismic and classical constraints during the computation of total $\chi^2$.
Finally, the different stellar parameters and their uncertainties  are obtained from the statistical mean and standard deviation of the posterior PDFs.

\section{Results and Discussion}
\label{results}

\begin{figure*}
	\includegraphics[width=\columnwidth]{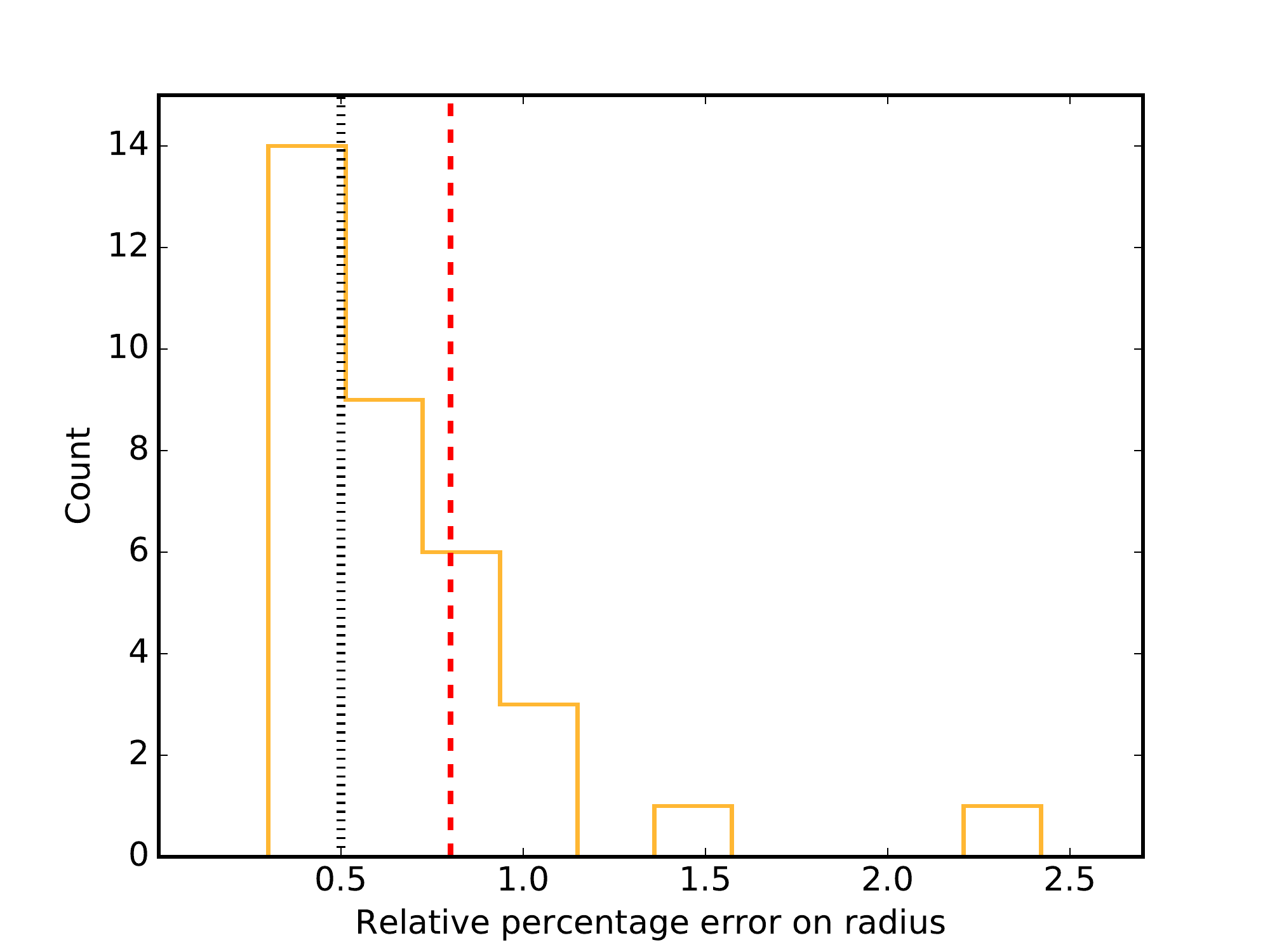}
	\includegraphics[width=\columnwidth]{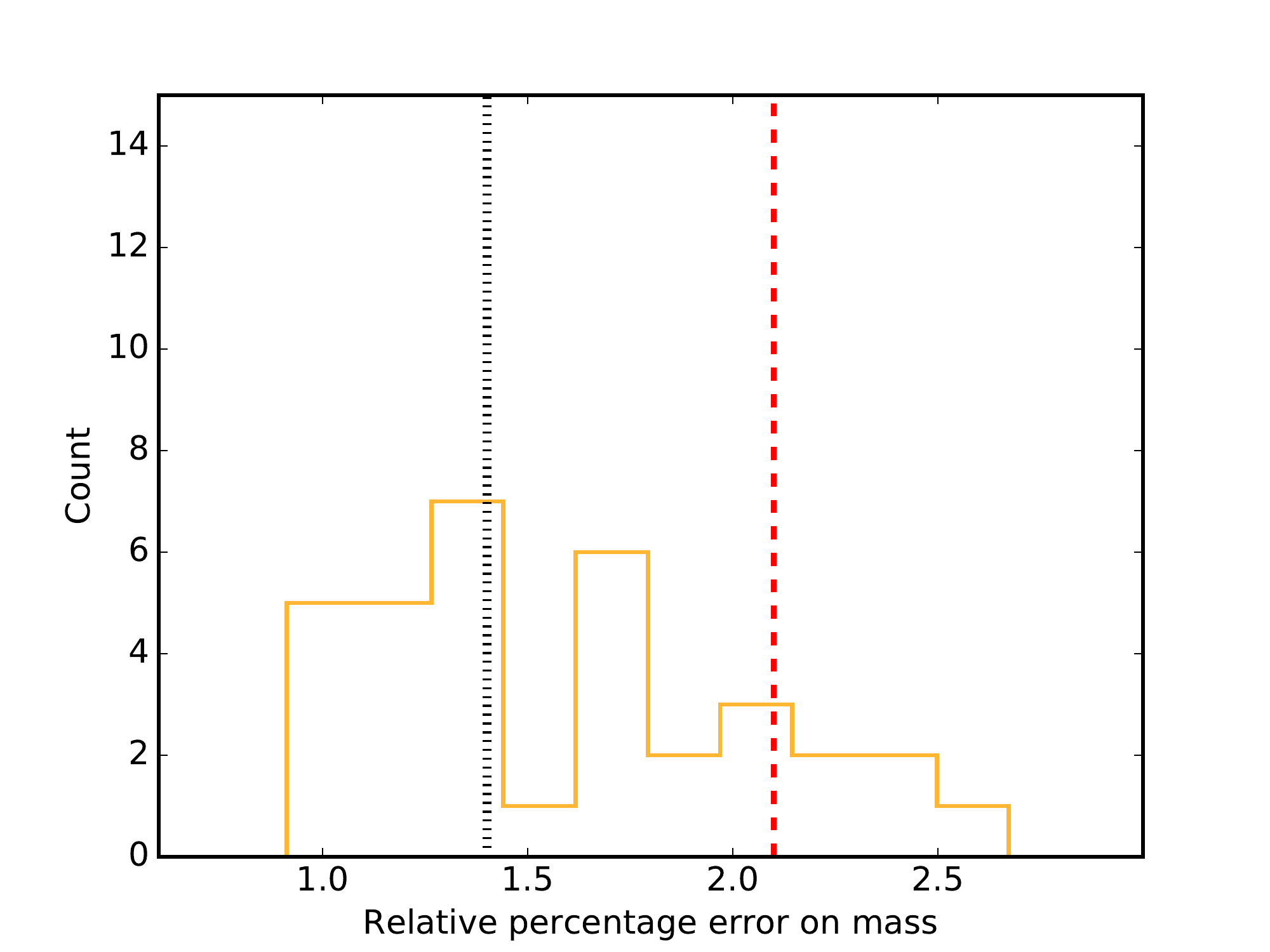}
    	\includegraphics[width=\columnwidth]{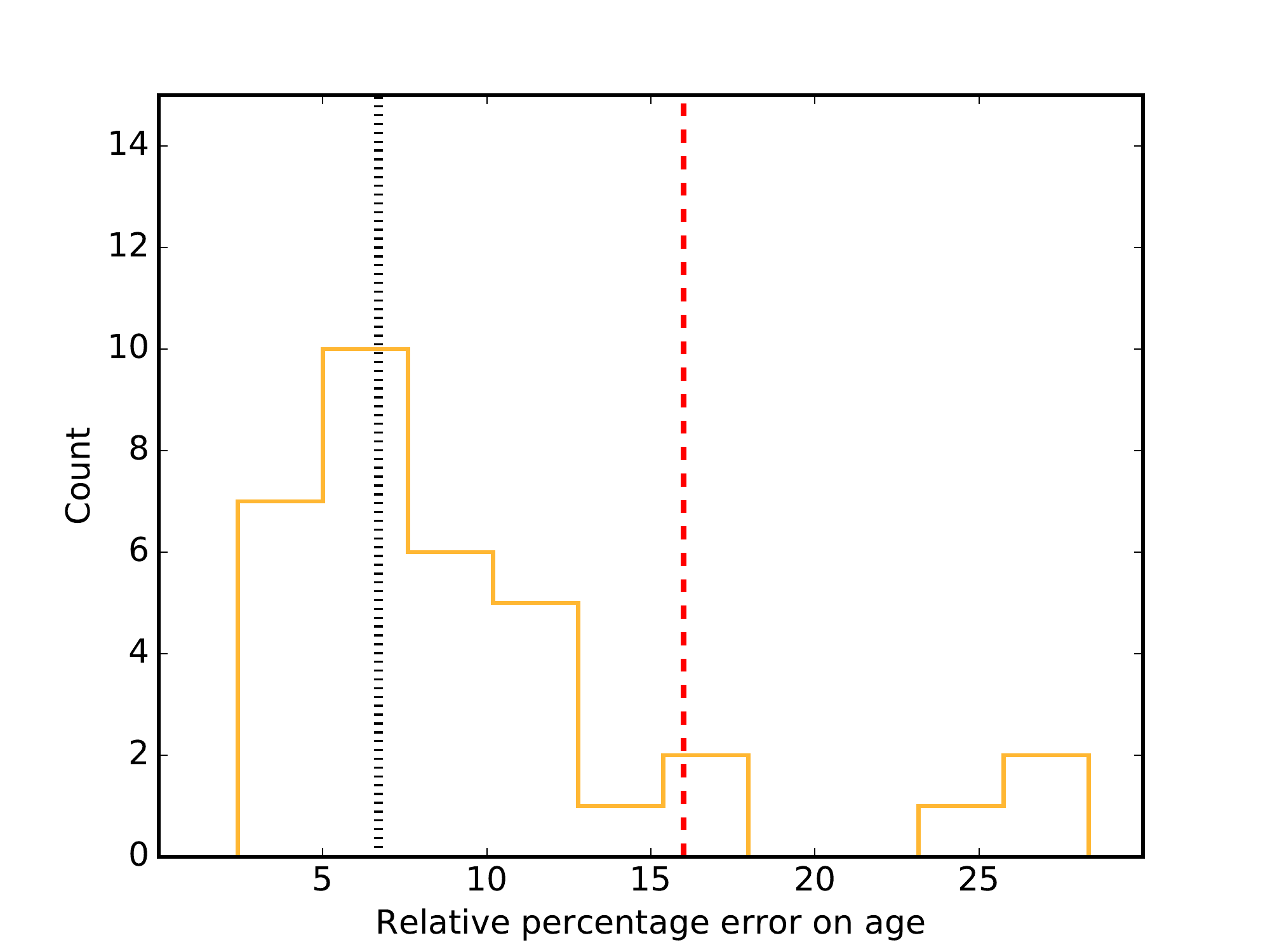}
            	\includegraphics[width=\columnwidth]{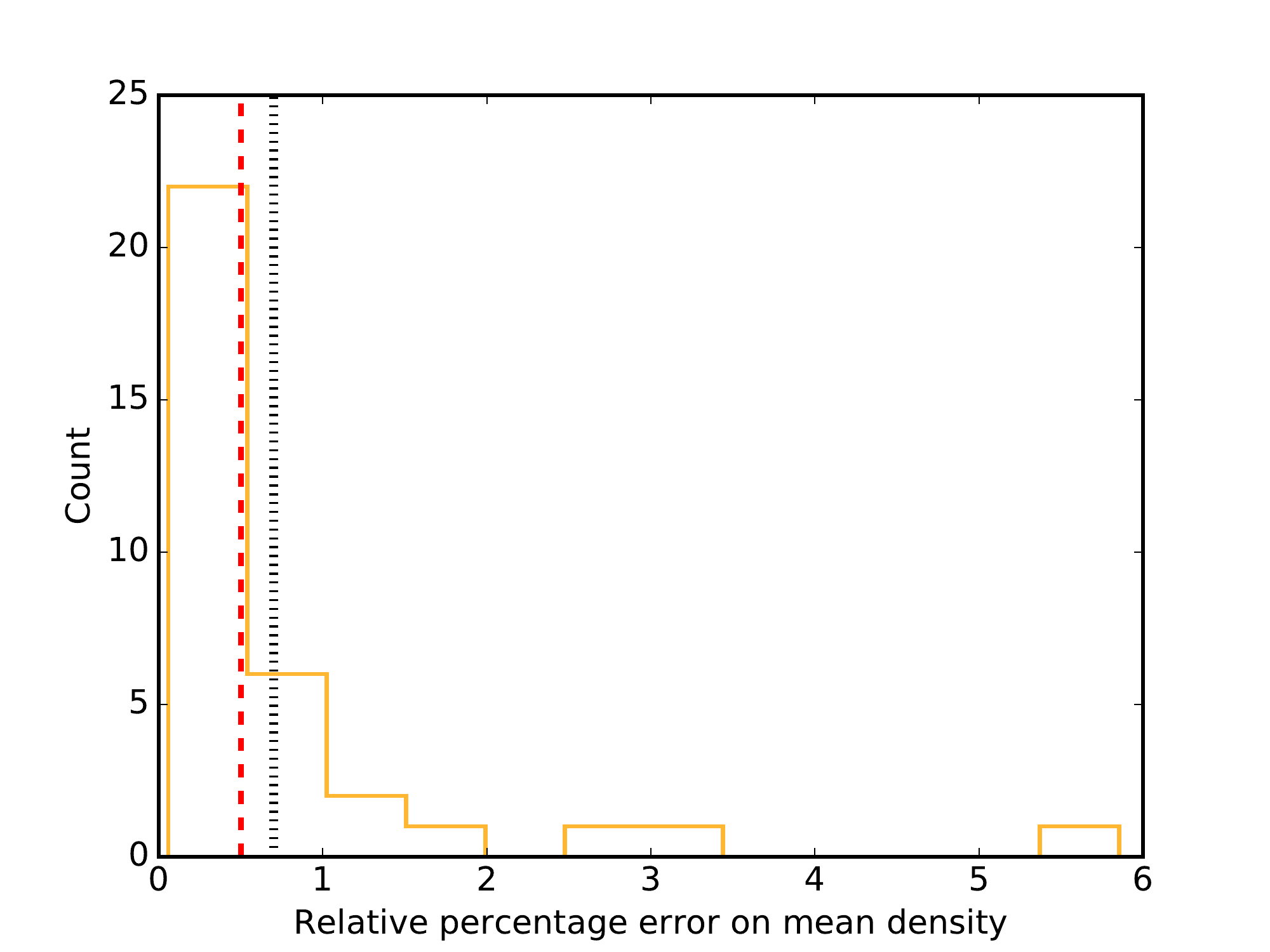}
      \caption{Statistical uncertainties and internal systematics. Histograms represent the distributions of statistical uncertainties when adopting the reference grid GS98sta. Black dotted lines represent internal systematic contributions from AGS09 (composition). Red dashed lines represent internal systematic contributions from GS98nod (diffusion).}
    \label{AGS09pre}
\end{figure*}

We discuss in detail the different input physics under investigation in Sects.~\ref{diffusion} and \ref{abundance}. It should be noted that some inputs cannot be examined separately since their modification requires changing other inputs. For instance, modifications in the solar metallicity mixture require setting the corresponding appropriate opacities. Therefore, in such cases, the systematics found  are from both sets of inputs. The two-term surface correction method by \cite{Ball} is used to obtain the results presented in Sects.~\ref{diffusion} and \ref{abundance}. In Sect.~\ref{freq}, we used the GS98sta grid in the analysis of the internal systematics arising from using different frequency correction methods. The percentage median statistical uncertainties obtained when using the reference grid GS98sta are 0.3\% in mean density, 0.6\% in radius, 1.6\% in mass, and 7.4\% in age (see Fig.~\ref{AGS09pre}).

\subsection{Diffusion}
\label{diffusion}
Atomic diffusion (or element diffusion) is a transport process that occurs in radiative regions of stars. 
It can be driven by temperature gradients (thermal diffusion), gravity or pressure gradients (gravitational settling), and composition gradients (chemical diffusion). Thermal diffusion and gravitational settling concentrate heavier elements towards the centre of the star \citep{Thoul}. These two processes
are opposed by composition gradients. Atomic diffusion is a less efficient process in convective regions since convection is a highly vigorous process that occurs on shorter timescales. Diffusion requires a quiet environment so that settling is not prevented by large-scale motions \citep{Chaboyer}. We switch on element diffusion in MESA, which includes chemical diffusion and gravitational settling \citep{Pax1}. MESA's diffusion module uses diffusion coefficients from \citet{Thoul} in order to solve Burger's equations when calculating particle diffusion and gravitational settling.

\begin{figure*}
	\includegraphics[width=\columnwidth]{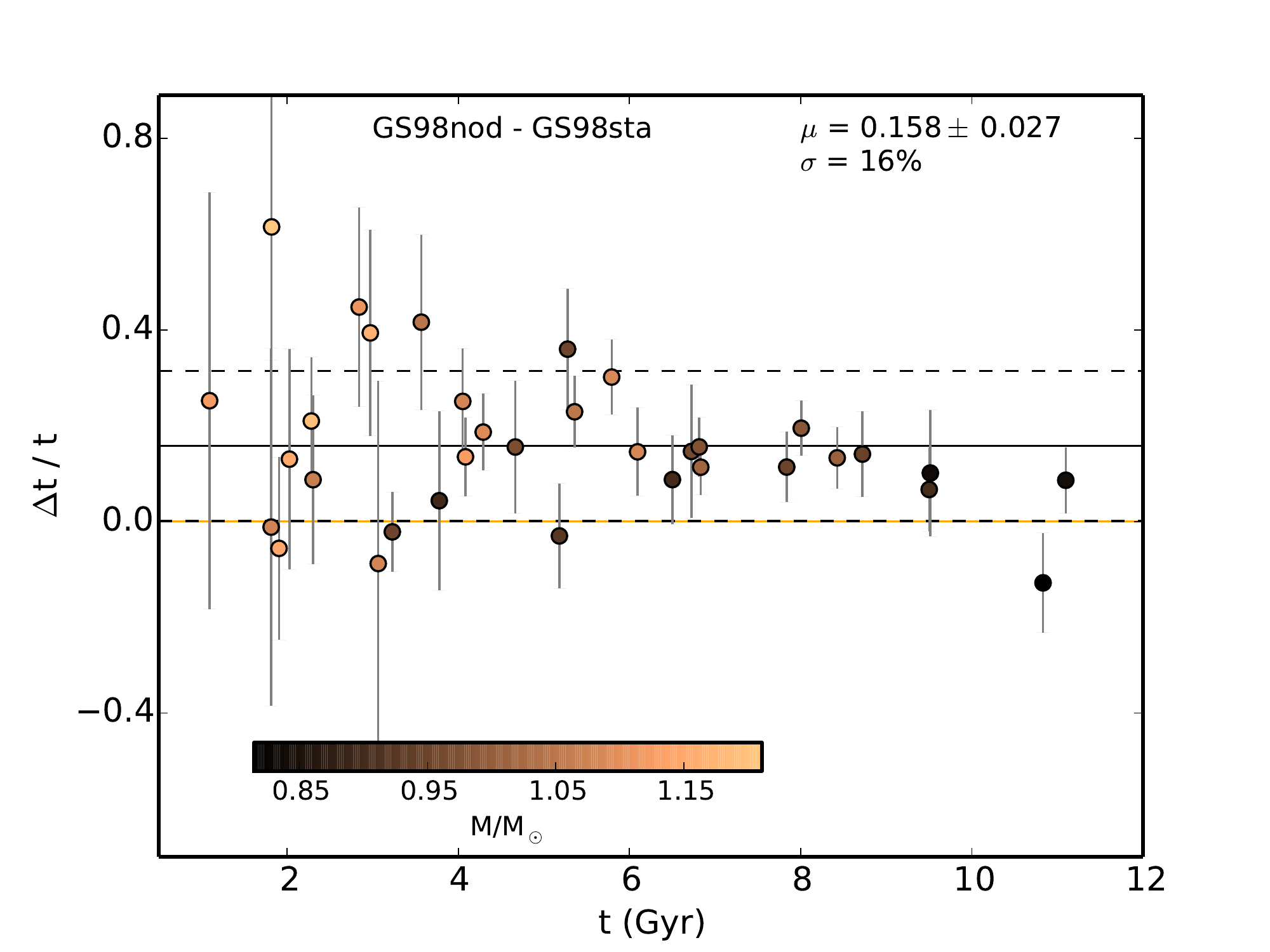}
    \includegraphics[width=\columnwidth]{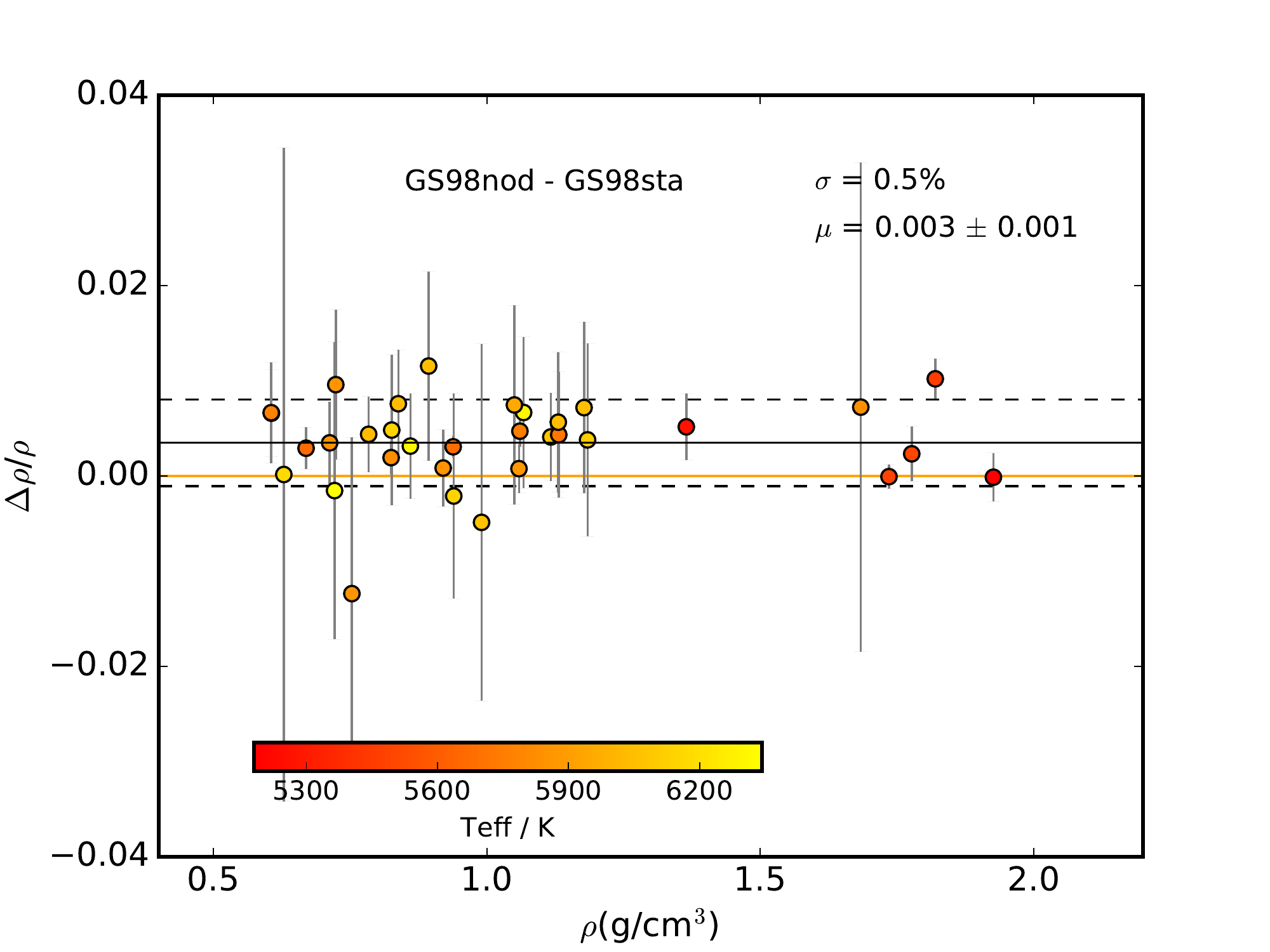}
      \caption{Fractional difference in age (left) and mean density (right) as a function of GS98sta stellar parameters. The colour-coding is with respect to stellar mass and effective temperature for the left and right panel, respectively. The solid black line indicates the bias ($\mu$), while the scatter ($\sigma$) is represented by the dashed lines. The zero level is represented by the solid orange line.}
    \label{agediff}
\end{figure*}

\begin{figure*} 
	\includegraphics[width=\columnwidth]{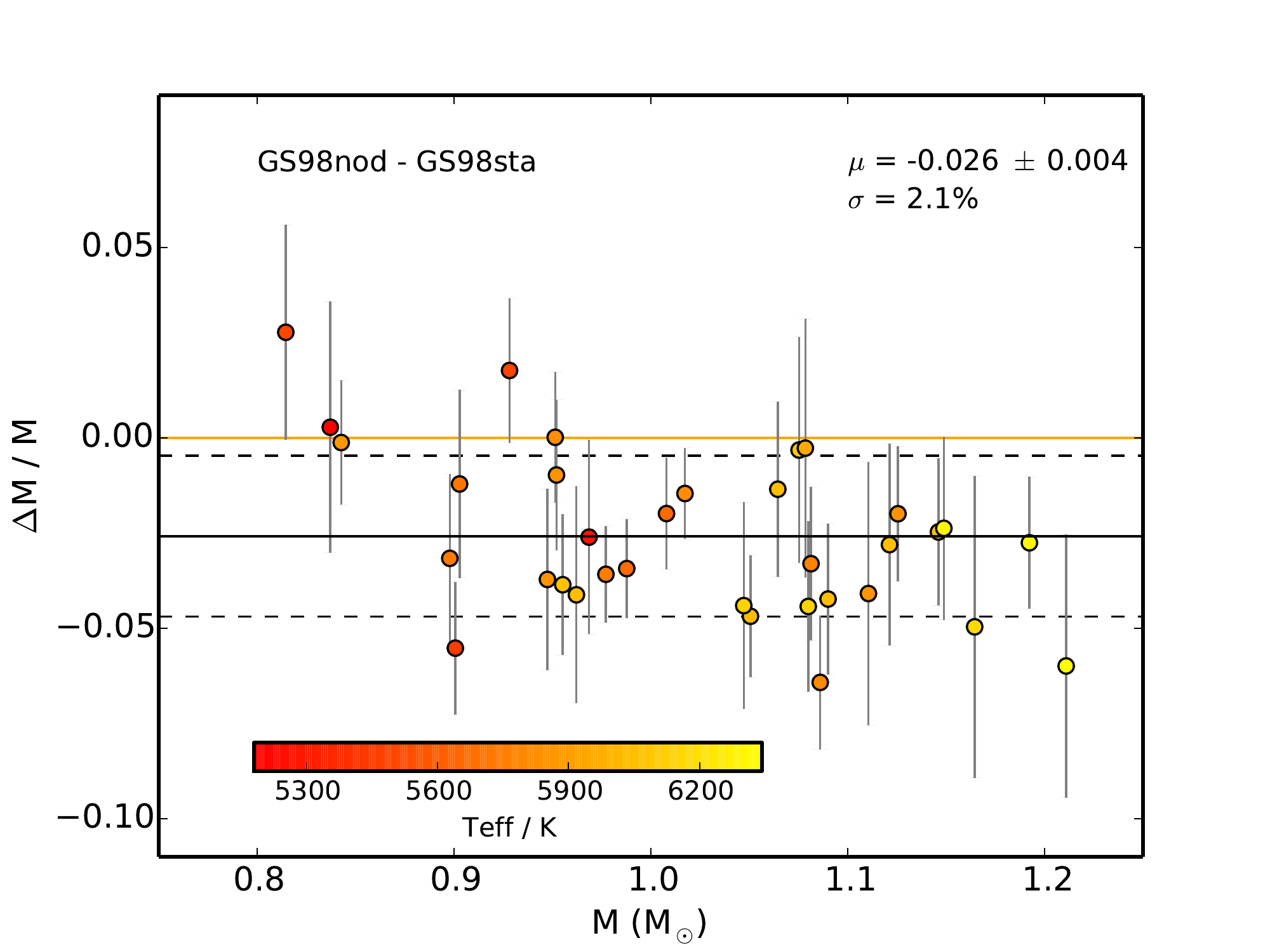}
	\includegraphics[width=\columnwidth]{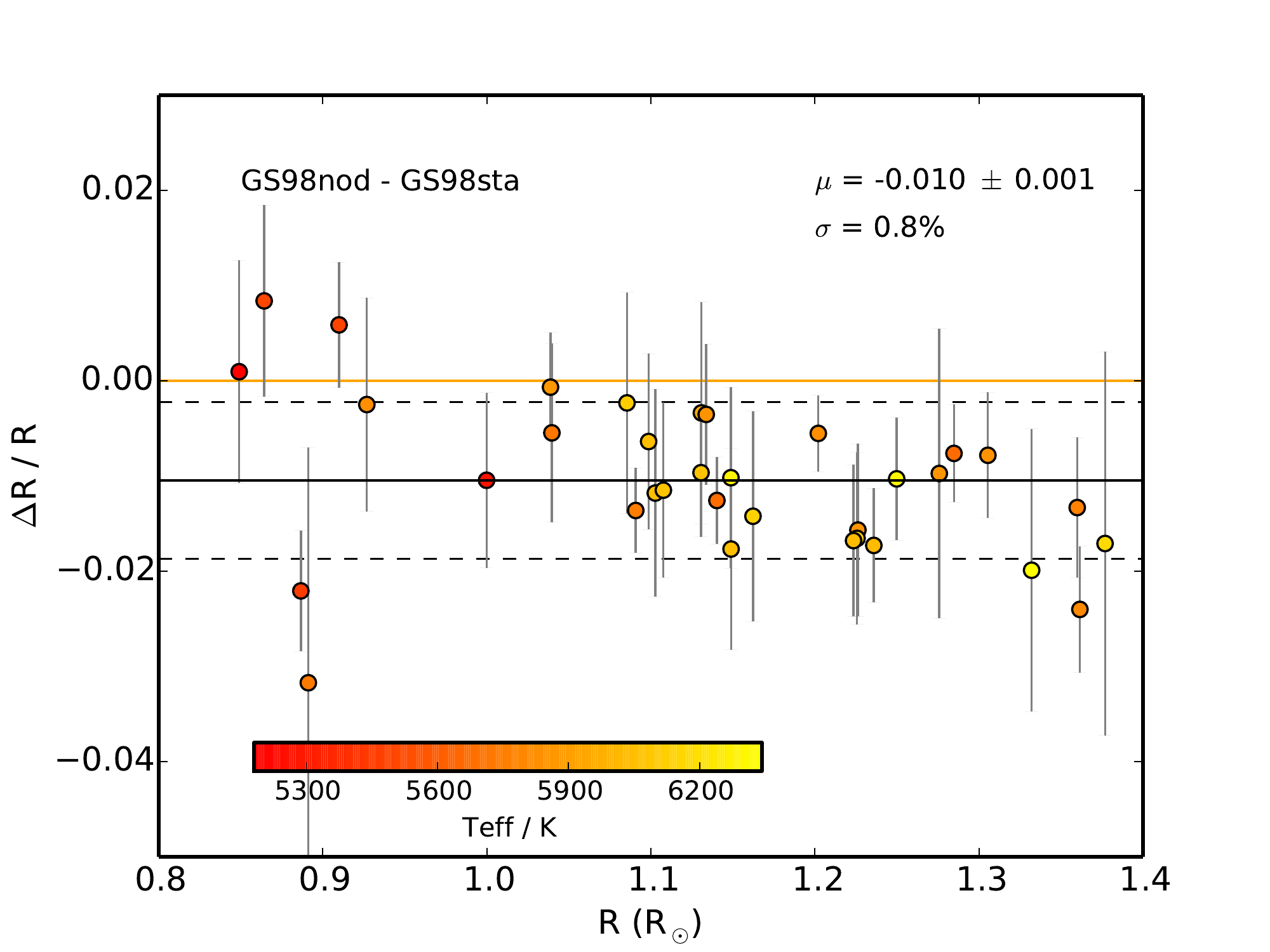}
      \caption{Fractional difference in mass (left) and radius (right) as a function of GS98sta stellar parameters. The zero level is represented by the solid orange line. The colour-coding is with respect to effective temperature. The solid black line indicates the bias ($\mu$), while the scatter ($\sigma$) is represented by the dashed lines.}
    \label{MR_diff}
\end{figure*}

Figure \ref{agediff} (left panel) shows that stellar ages derived using GS98nod are systematically larger than those derived using GS98sta. 
In our optimization process, we use the effective temperature and metallicity as classical observables. Since element diffusion changes surface element abundances, our GS98sta grid will need to have best fitting models with higher initial metal mass fractions so that the surface metal mass fractions can be comparable to the observed values at the stars' current ages.  
This implies the opacity in the cores of these models will be higher throughout their evolution, 
compared to the case of no diffusion. To avoid the associated decrease in luminosity, which is indirectly constrained by the effective temperature and seismic data, the best fit models need to have higher mass, justifying their younger age. This is confirmed in  Fig.~\ref{MR_diff} (left panel) where we show the fractional differences in the stellar mass. In turn, the strong constraints on the mean density imposed by the seismic data lead to an increase in the radius of the best fit models in our GS98sta grid, as shown in  Fig.~\ref{MR_diff} (right panel). In summary, in order to satisfy the observables, models with diffusion need to have higher masses, hence also higher radii and younger ages.

We explore internal systematics by comparing models with and without diffusion, resulting in 0.5\%, 0.8\%, 2.1\%, and 16\% in mean density, radius, mass, and age, respectively (see Figs.~\ref{agediff} and \ref{MR_diff}). The internal systematics in density, mass, and radius are comparable to their statistical uncertainties, while the internal systematics in age are significantly larger than the statistical uncertainties (See Fig.~\ref{AGS09pre}). 

Furthermore, we inspect the data presented in both Fig.~\ref{agediff} and Fig.~\ref{MR_diff} for the presence of linear correlations. We employ a correlation coefficient analysis developed in a Bayesian framework \citep{Figueira}. Using it we can estimate the posterior probability distribution of the correlation coefficient. Table~\ref{correlation} shows the results from such Bayesian test. At the 95\% confidence level, there is a negative linear trend present for the radius and mass, with a hint of a similar trend being present for the age. 
\begin{table}
\centering 
\caption{Statistical summary of the posterior probability distribution of the correlation coefficient. C.I.~denotes the confidence interval.}
\begin{tabular}{l|c|c|r}        %
\hline 
Parameter	& Mean		& Standard Deviation   &    95\% C.I.   \\
\hline
Mass  	    & -0.433 	&  0.135   &    [-0.69, -0.168] 	\\
Radius 	    & -0.342    &  0.142   &  	[-0.599, -0.049]		  \\
Age			& -0.247	&  0.152   &    [-0.535, 0.047]          \\
\hline                                   
\end{tabular}
\label{correlation}
\end{table}
The difference between the initial and current metal mass fraction, $Z$, of the best-fitting models of GS98sta increases with increasing stellar mass. This suggests that gravitational settling has a larger impact on higher mass stars, explaining the trend with mass seen in Fig.~\ref{MR_diff}.

\subsection{Composition}
\label{abundance}

Solar metallicity mixtures are one of the most important ingredients in stellar modelling. Here we focus on the most commonly used solar metallicity mixtures in constructing standard solar models, namely, those from \citet{Grevesse} and \citet{Asplund}.
We define [Fe/H] in all our calculations as
\begin{equation}
\mbox{[Fe/H]} = \log\left( \frac{Z_{\rm surface}}{X_{\rm surface}}\right)_{\rm star} - ~\log\left( \frac{Z_{\rm surface}}{X_{\rm surface}}\right)_\odot ~~~ ,
\end{equation}
where $X_{\rm surface}$ and $Z_{\rm surface}$ refer to the surface hydrogen and heavy element mass fractions, respectively. We used solar $Z_{\rm surface}$ values of 0.0134 and 0.0169 based on \citeauthor{Asplund} and \citeauthor{Grevesse}, respectively.
In general, the solar metallicity mixture from AGS09 are lower than those from GS98sta. 
\cite{Basu} demonstrate that the uncertainty in solar metallicity mixture result in differences in the sound speed in the stellar interiors. 

\begin{figure*}
	\includegraphics[width=\columnwidth]{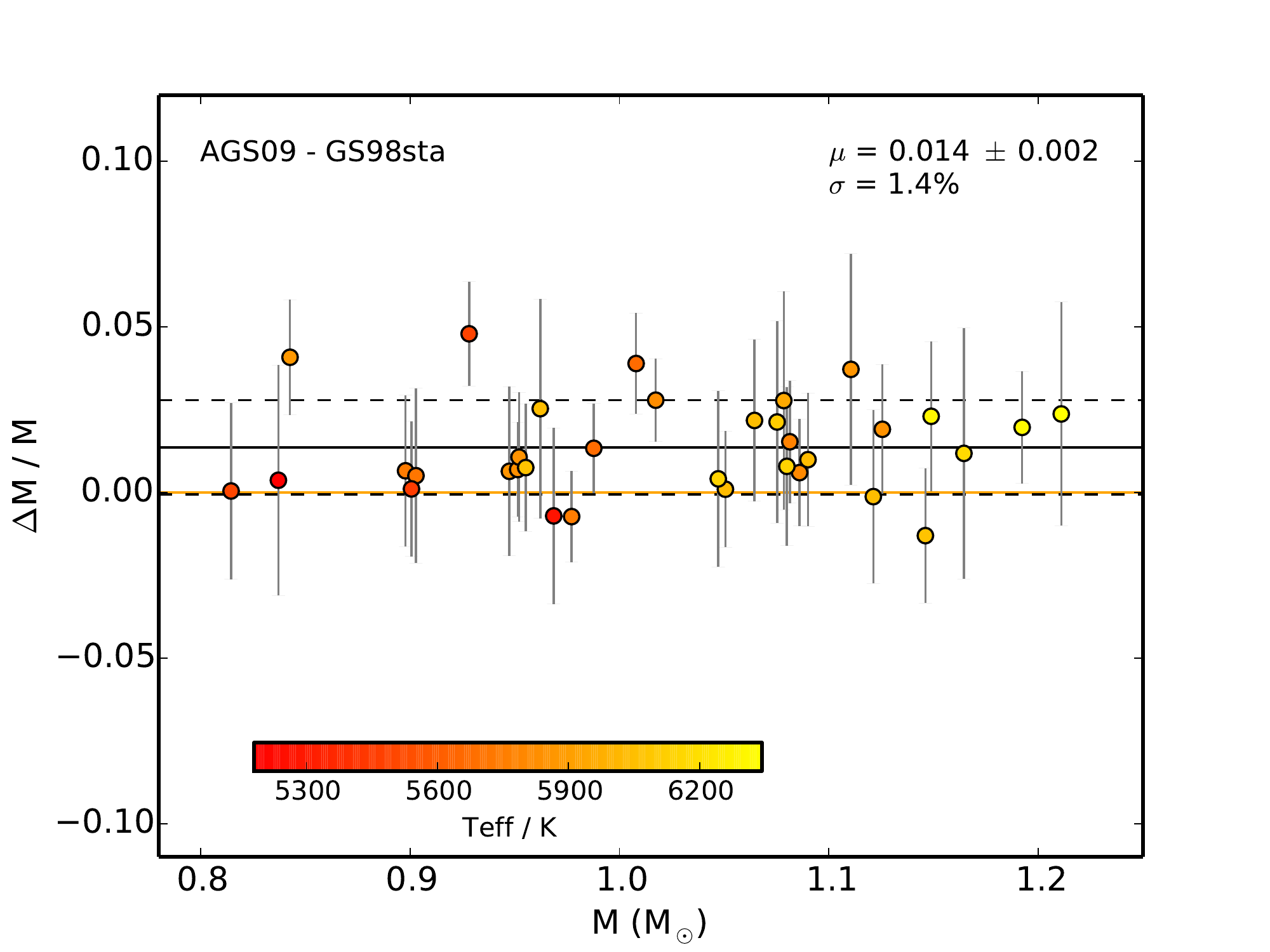}
	\includegraphics[width=\columnwidth]{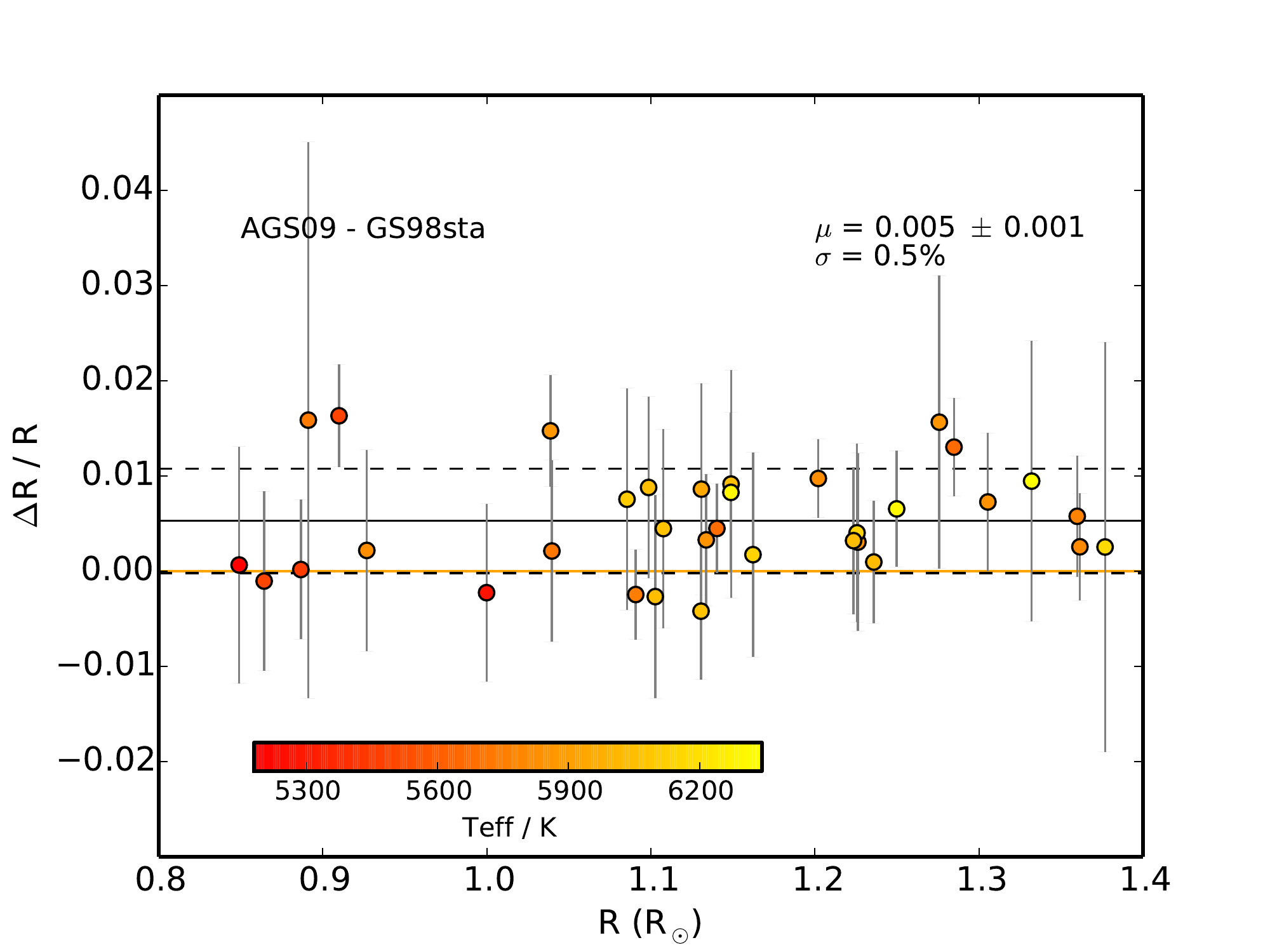}
      \caption{Fractional difference in mass (left) and radius (right) as a function of GS98sta stellar parameters. The zero level is represented by the solid orange line. The colour-coding is with respect to effective temperature. The solid black line indicates the bias ($\mu$), while the scatter ($\sigma$) is represented by the dashed lines.}
    \label{MRcomp}
\end{figure*}

We assess the internal systematics arising from the uncertainty in solar metallicity mixture.
\begin{figure*}
	\includegraphics[width=\columnwidth]{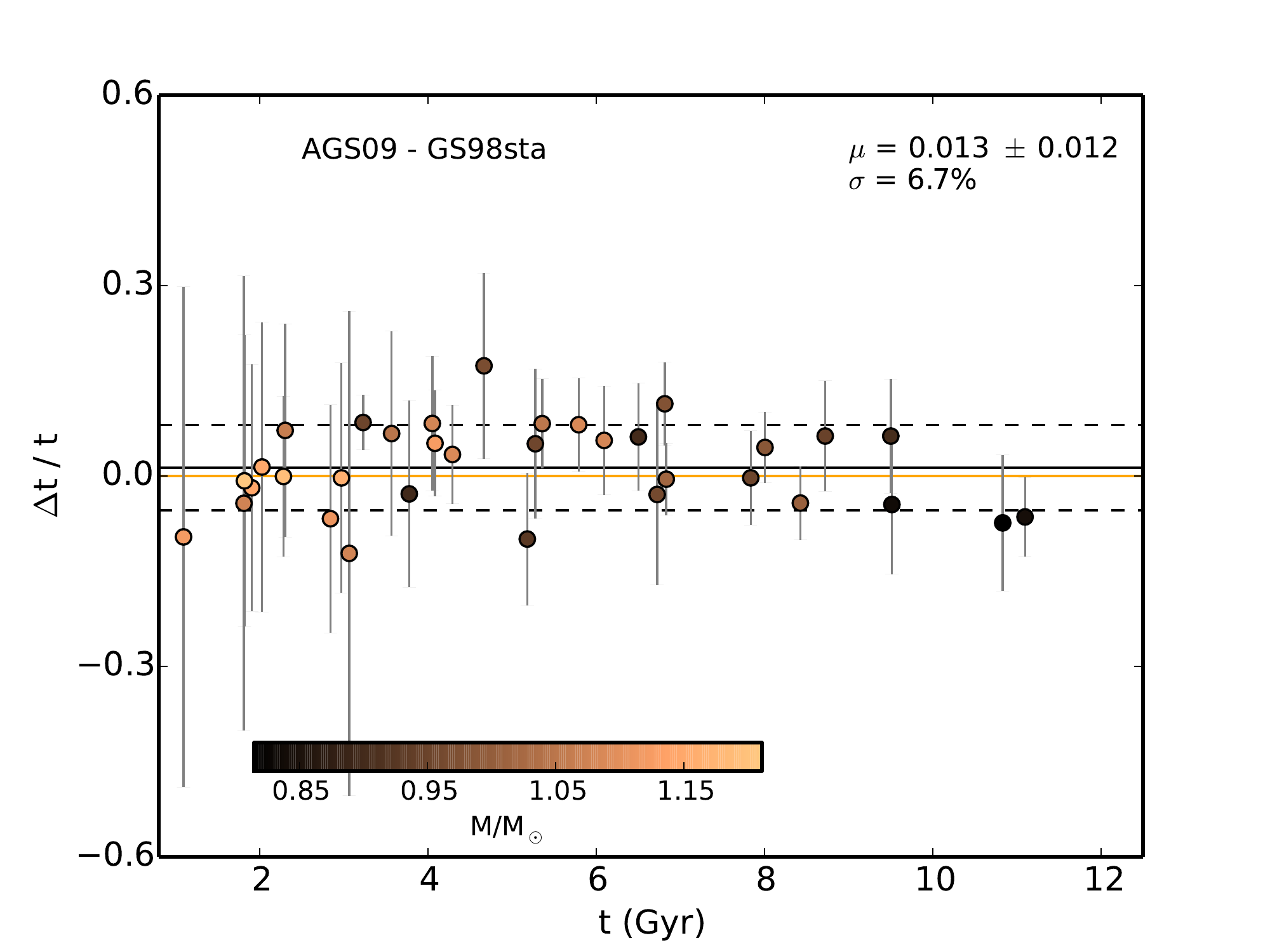}
    	\includegraphics[width=\columnwidth]{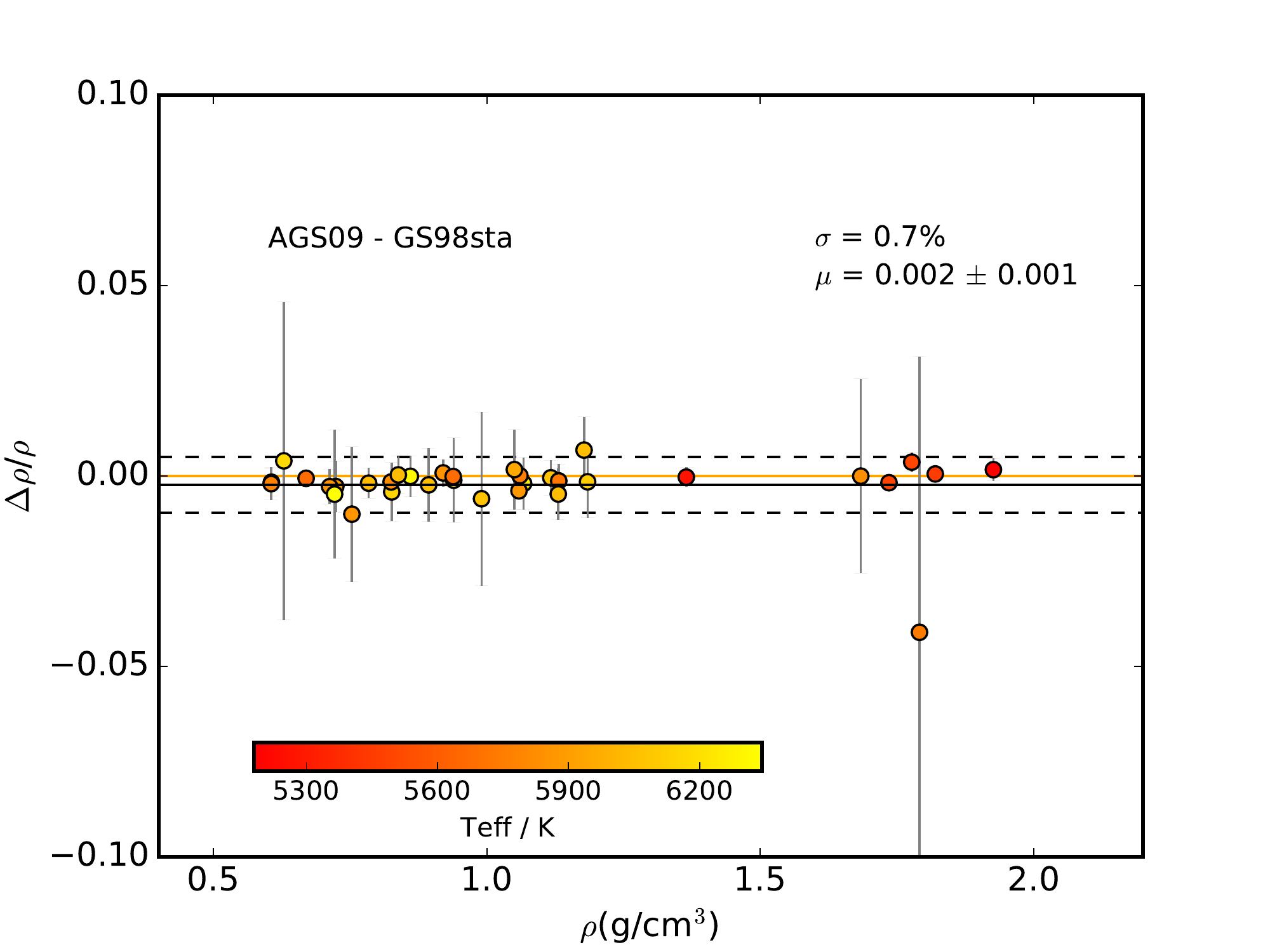}
      \caption{Fractional difference in age (left) and mean density (right) as a function of GS98sta stellar parameters. The colour-coding is with respect to stellar mass. The zero level is represented by the solid orange line. The solid black line indicates the bias ($\mu$), while the scatter ($\sigma$) is represented by the dashed lines.}
    \label{Agecomp}
\end{figure*}
We find a good agreement in both the derived ages and densities with internal systematics of 6.7\% (with a bias of 0.013 $\pm$ 0.012) and 0.7\% (with a bias of 0.002 $\pm$ 0.001), respectively (see Fig.~\ref{Agecomp}). Internal systematics in stellar ages are somewhat smaller than the statistical uncertainties (median of 7.4\%) as shown in Fig.~\ref{AGS09pre}. 
Furthermore, we find internal systematics of 0.5\% in radius and 1.4\% in mass (see Fig.~\ref{MRcomp}). 
\citet{Aguirre} found systematic contributions arising from the uncertainty in solar metallicity mixture to be 0.3\% in mean density and radius, 0.6\% in mass and 3.3\% in age. The internal systematics found in this work are approximately twice as large as those found by \citet{Aguirre}. The most probable cause for this discrepancy is in the treatment of the mixing length parameter ($\alpha_{\rm mlt}$). In this paper, we set $\alpha_{\rm mlt}$ as a free variable in all of our grids (see Sect.~\ref{grid1}), while \citet{Aguirre} used solar calibrated $\alpha_{\rm mlt}$ values.  We note that the uncertainty in solar metallicity mixture will cause variations in solar calibrated values of $\alpha_{\rm mlt}$. Systematic uncertainties arising from grids varying in the treatment of $\alpha_{\rm mlt}$ were found to be 0.7\%, 0.6\%, 2.2\%, and 9.0\% in mean density, radius, mass, and age, respectively \citep{Aguirre}. In addition, based on a grid of 3D convection simulations, $\alpha_{\rm mlt}$ has been shown to span a range of values on the  main-sequence \citep{Nordlund}. It is therefore advisable to use a range of $\alpha_{\rm mlt}$ values when constructing stellar grids for Sun-like stars.

\subsection{Surface correction}
\label{freq}
To overcome the well-known systematic differences between model and observed oscillation frequencies (Sect.~\ref{grid}),
several surface correction methods have been put forward (see Table \ref{3}).
The function, $f$, used in the different surface correction methods is given by
\begin{equation}
f = \frac{\nu}{\nu_0} ~~,
\end{equation}
where $\nu$ is the mode oscillation frequency and $\nu_0$ a reference frequency. 
When using surface correction method proposed by \citet{Sonoi}, the value of $\nu_0$ is determined in AIMS by means of the scaling relation \citep{Brown,Kjeldsen2}:
\begin{equation}
\nu_0 = \left(\frac{M}{\mbox{M}_\odot}\right) \left(\frac{R}{\mbox{R}_\odot}\right)^{-2} \left(\frac{T_{\rm eff } }{T_{{\rm eff},\odot}} \right)^{-1/2} \nu_{{\rm max},\odot}~~.
\end{equation}
The adopted solar values are $\nu_{{\rm max},\odot} = 3104.0\:{\rm \mu Hz}$, ${\rm R}_\odot = 6.9599\times10^{10}\:{\rm cm}$, ${\rm M}_\odot = 1.98919\times10^{33}\:{\rm g}$, and $T_{\rm eff,\odot} = 5777.0\:{\rm K}$ \citep{Allen,Mamajek,mos}.
\begin{table}
\centering 
\caption{ Summary of the different surface correction methods. $a$ and $b$ are best-fit parameters (see text for details), $f$ is a function that depends on the mode frequency, $I$ is the normalised  mode inertia, and $\delta\nu$ is the offset between observed and model frequencies. }
\begin{tabular}{lccr}        
\hline 
Name 	&	Functions		        &  b - value               & Reference	\\
\hline
KJ 	&   $\delta\nu = af^b$			&  4.9	                        & \cite{Kjeld} \\
BG1 	&   $\delta\nu = a f^3 / I$  &  -         	& \cite{Ball} \\
BG2	&   $\delta\nu = (af^{-1} + bf^{3}) /I$ &  -  	& \cite{Ball}  \\
Sonoi   &   $\delta\nu = a\left(1 -\frac{1}{1+f^b}\right)$&  4.0       &	\cite{Sonoi}	  \\
\hline                                   
\end{tabular}
\label{3}
\end{table}
For other surface corrections, we used
\begin{equation}
\nu_0 = \frac{1}{2\pi}~\sqrt[]{\frac{GM}{R}} ~~,
\end{equation}
where $G$ and $R$ are the gravitational constant and model radius, respectively. This does not make a difference on the correction, but only affects the magnitude of the fitting coefficients (i.e., $a$ and $b$).
Some striking differences between the different surface correction methods are worth mentioning.
\cite{Kjeld} proposed that the offset depends on a power of the mode frequency, whose exponent, $b$, they determined to be $b = 4.9$ (calibrated with respect to solar data). The same value has subsequently been adopted in the study of other stars \citep{Thompson,Brand,Eylen,Gruberbauer}. Using Canuto-Goldman-Mazzitelli (CGM)  modelling of convection, \citet{Deheuvels2014} determined a value of $b$ to be $4.25$ and adopted it in generating models using the Cesam2k evolutionary code.
We adopted 
the former value of $b$ in AIMS when using the surface correction of \cite{Kjeld}, as shown in Table~\ref{3}. 
\citet{Sonoi} established that the power-law function proposed by \citet{Kjeld} is not satisfactory in fitting the high-frequency range. This was attributed to the nature of the profile of the frequency difference, $\delta\nu$, which becomes less steep as the frequency increases beyond $\nu_{\rm max}$.
\citet{Sonoi} thus proposed a Lorentzian function that they found to better fit the profile of the frequency difference across the whole frequency range. It should be noted that the modified Lorentzian function proposed by \cite{Sonoi} reduces to the equation proposed by \cite{Kjeld} when $f \ll 1$. 
Furthermore, the scaling factor, $r$, related to the mean density and proposed by \cite{Kjeld}, is not used in AIMS. The risk with rescaling the model is that one 
will need to change a number of variables (e.g. how is heat transport affected). It may be possible to rescale the acoustic variables consistently, but other 
variables may not remain consistent. Hence, for this reason the $r$-scaling is not implemented in AIMS. If there is a mismatch, then AIMS looks for another model where the mean 
density is closer, rather than trying to rescale the model to the right mean density. 
\cite{Ball} proposed two functions (BG1 and BG2, see Table~\ref{3}), both taking into account the mode inertia, $I$. This was based on findings of  \citet{Gough1990} and \citet{Goldreich} who argued that perturbations caused by a magnetic field would cause changes proportional to $\nu^3/I$ and a 
change in the description of convection is expected to cause changes proportional to $\nu^{-1}/I$. BG1 takes into account only the cubic term while BG2 combines both terms (see Table~\ref{3}). \\
We note that we used the same set of observed frequencies for each star when applying the different surface correction methods during the optimisation process. This is because we aimed at carrying out a uniform analysis for all the different surface correction methods.

We explore the internal systematics from adopting the different surface correction options by comparing the model parameters derived in each case with those obtained when using frequency ratios. 
Frequency ratios have been shown to be less affected by the poorly modelled surface layers and this permits direct comparison of observed oscillation frequencies with the theoretical oscillations frequencies without applying any surface correction routine (\citealt{Rox, Silva,Aguirre,Aguirre1}).  Unfortunately, some information about the star is lost when one uses frequency ratios. For instance, since the stellar mean density scales with the frequencies, taking frequency ratios results into factoring out the mean density and thus making frequency ratios less sensitive to the mean density compared to direct comparison with oscillation frequencies \citep{Rox}. Despite this, frequency ratios have been reported to constrain stellar interiors, resulting in more precise asteroseismic stellar ages compared to the use of individual frequencies.
In AIMS,
we specified the frequency ratios $r_{10}$, $r_{01}$, and $r_{02}$ as \citep{Rox}:
\begin{equation}
 r_{01}(n) = \frac{d_{01}(n)}{\Delta\nu_1(n)}~,~~~ r_{10}(n) = \frac{d_{10}(n)}{\Delta\nu_0(n+1)}~~,
 \end{equation}

\begin{equation}
 r_{02}(n) = \frac{d_{02}(n)}{\Delta\nu_1(n)}~,
\end{equation}
where $\Delta\nu_l (n) = \nu_{n,l}-\nu_{n-1,l}$ is the large frequency separation, $n$ is the mode radial order, $l$ is the mode harmonic degree, and $d_{02}(n)=\nu_{n,0}-\nu_{n-1,2}$ is the small frequency separation. Moreover, $d_{01}(n)$ and
$d_{10}(n)$ are defined as:
\begin{equation}
 d_{01}(n)=\frac{1}{8}(\nu_{n-1,0} - 4\nu_{n-1,1} + 6\nu_{n,0} - 4\nu_{n,1} + \nu_{n+1,0})~~,
\end{equation}

\begin{equation}
 d_{10}(n)= -\frac{1}{8}(\nu_{n-1,1} - 4\nu_{n,0} + 6\nu_{n,1} - 4\nu_{n+1,0} + \nu_{n+1,1})~~.
\end{equation}

\begin{figure*}
    \centering
    \includegraphics[width=15cm, height=10cm]{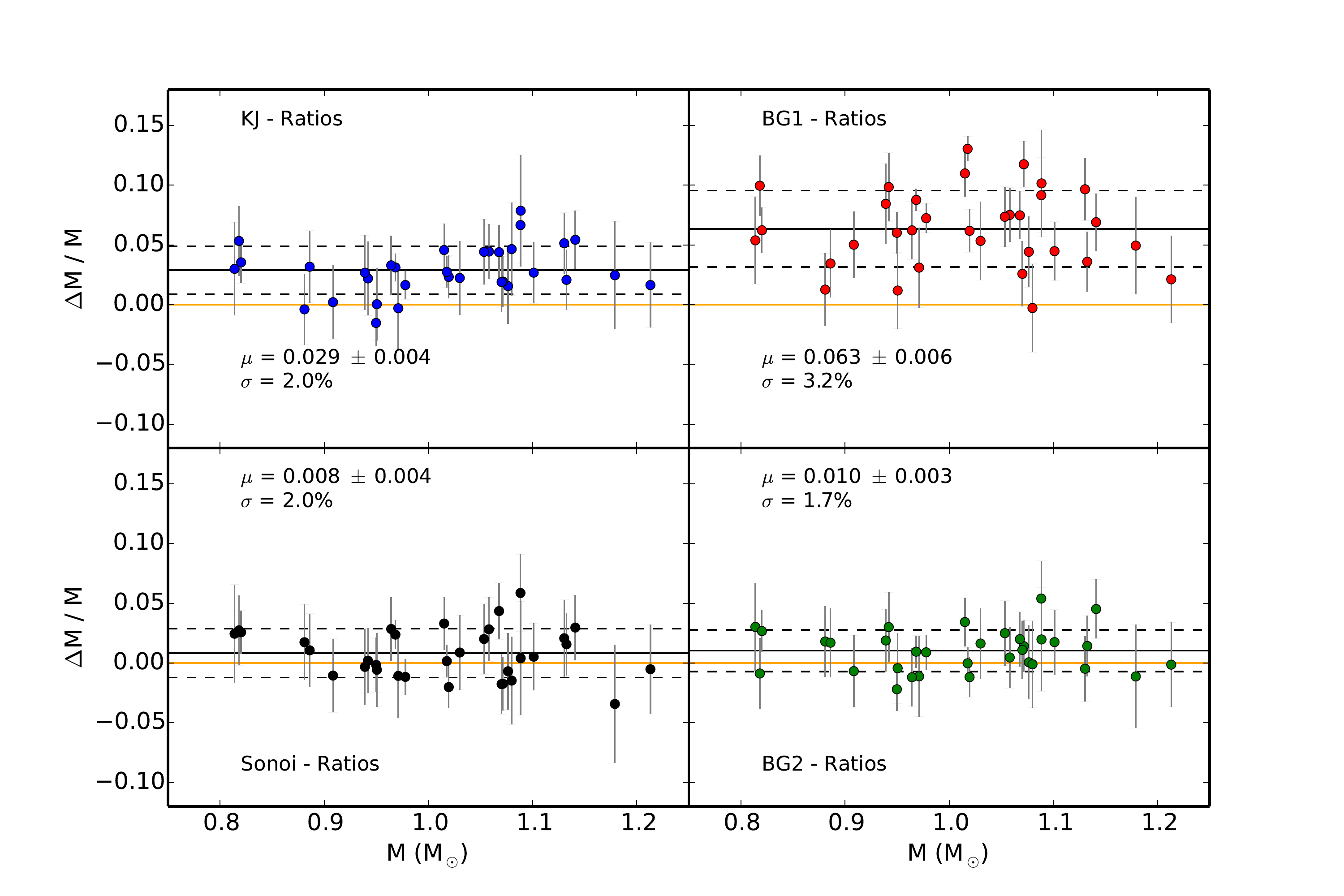}
    \caption{Fractional difference in mass as a function of GS98sta stellar masses. The zero level is represented by the solid orange line. The solid black line indicates the bias ($\mu$), while the scatter ($\sigma$) is represented by the dashed lines.}
    \label{massfreq}
\end{figure*}

\begin{figure*}
    \centering
    \includegraphics[width=15cm, height=10cm]{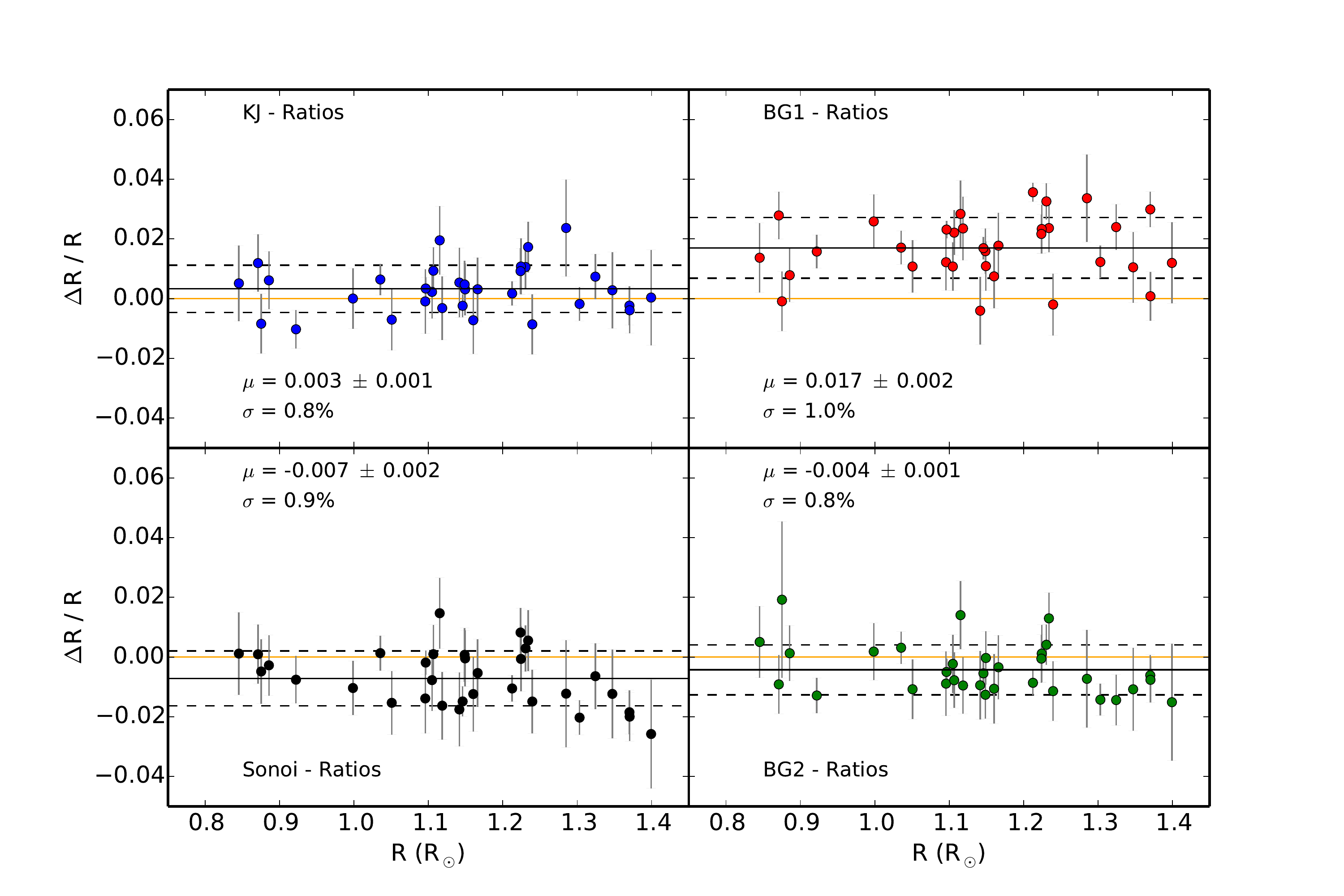}
    \caption{Fractional difference in radius as a function of GS98sta stellar radii. 
    The solid black line indicates the bias ($\mu$), while the scatter ($\sigma$) is represented by the dashed lines. The zero level is represented by the solid orange line.}
    \label{radiusfreq}
\end{figure*}

Hereafter, the surface correction method by \cite{Kjeld} is denoted by KJ, \cite{Sonoi} as Sonoi, \cite{Ball} one-term correction as BG1, and  \cite{Ball} two-term correction as BG2 (See Table \ref{3}). Sonoi and BG2 lead to smaller internal systematics in mass: 2.0\% and 1.7\%, respectively (see Fig.~\ref{massfreq}). We find that the masses are overestimated when employing the corrections by KJ and BG1. KJ also yields internal systematics of 2.0\% in mass, albeit affected by a larger bias of 0.029 $\pm$ 0.004. All surface correction routines yield similar internal systematics in radius, with BG1 leading to the largest bias (see Fig.~\ref{radiusfreq}). Both KJ and BG2 produce internal systematics in radius of 0.8\%, while Sonoi yields 0.9\%.
\begin{figure*}
    \centering
    \includegraphics[width=15cm, height=10cm]{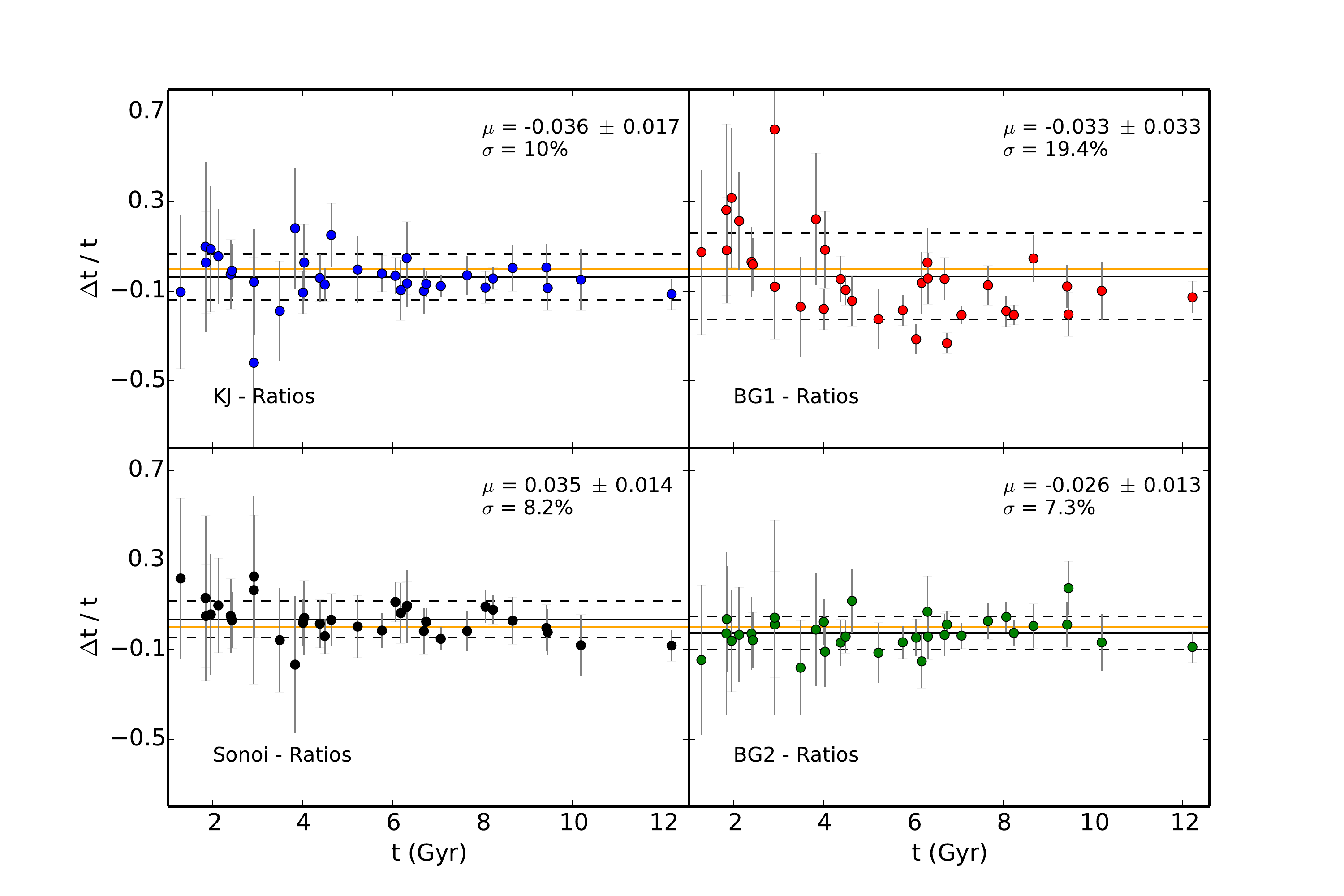}
    \caption{Fractional difference in age as a function of GS98sta stellar ages. The zero level is represented by the solid orange line. The solid black line indicates the bias ($\mu$), while the scatter ($\sigma$) is represented by the dashed lines.}
    \label{agefreq}
\end{figure*}
\begin{figure*}
    \centering
    \includegraphics[width=15cm, height=10cm]{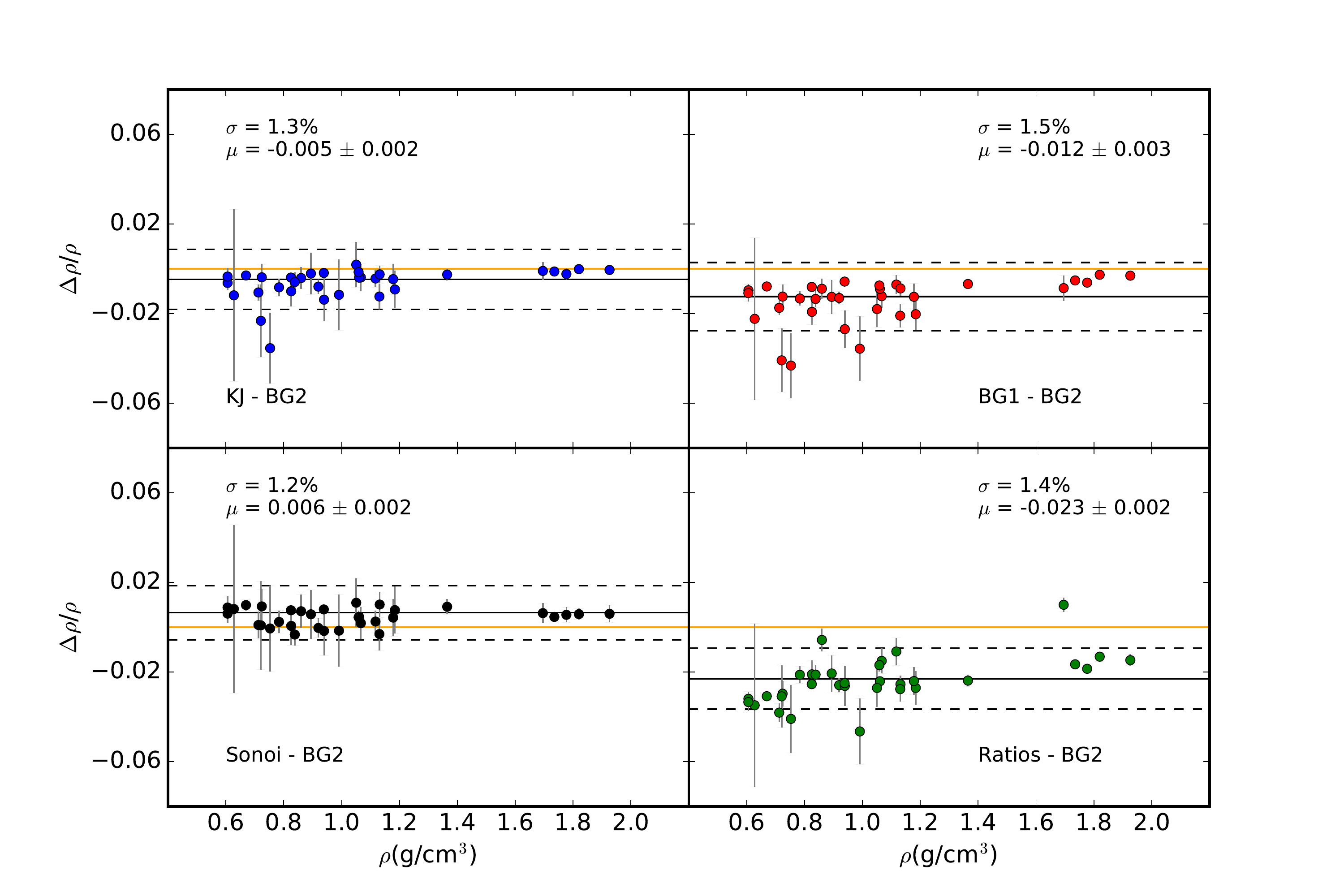}
    \caption{Fractional difference in stellar mean density as a function of GS98sta stellar mean density. The zero level is represented by the solid orange line. The solid black line indicates the bias ($\mu$), while the scatter ($\sigma$) is represented by the dashed lines.}
    \label{densityfreq}
\end{figure*}
Figure \ref{agefreq} shows that  Sonoi and BG2 produce smaller internal systematics in age: 8.2\% and 7.2\%, respectively. KJ results in internal systematics of 10\%, while BG1 yields the largest internal systematics (19.4\%).

BG2 yields the least median reduced $\chi^2_\nu$ in the model-to-observed frequency differences of 5.661. Sonoi and KJ yield comparable median reduced $\chi^2_\nu$ of 15.846 and 15.958, respectively. BG1 gives median reduced $\chi^2_\nu$ of 27.800. This in turn explains why the BG2 lead to the least internal systematics followed by Sonoi.

In Fig.~\ref{densityfreq}, we consider BG2 results as the reference in the comparison with the different surface correction methods. This is because from Figs.~\ref{massfreq}, \ref{radiusfreq}, and \ref{agefreq}, BG2 yields the least internal systematics in  mass, radius, and age, respectively. In addition, frequency ratios will provide a poor description of the near-surface layers. It can clearly be seen in Fig.~\ref{densityfreq} that some information about the mean density is lost when one uses frequency ratios.
Since we used uncorrected model frequencies to compute the frequency ratios as well as the large frequency separation, the later includes a significant contribution from the surface effect. Assuming that the large frequency separation for the best fitting model matched reasonably well with the observed separation, the "true" large separation for the model (excluding the surface effect contribution) is significantly underestimated, and hence the corresponding mean density. 
This is consistent with previous findings by \cite{Aguirre1}. The internal systematics on the mean density arising from varying the surface correction methods are found to be less than 1.5\%.

We note that when one compares the observed frequencies with the uncorrected theoretical frequencies of the best-fit models obtained using the different surface correction methods, surface corrections should be expected to tend to zero only for low enough frequencies. This is because, assuming the shape of the surface correction does not match with the differences between the observed and ``true model'' frequencies, then the best fitting model obtained by 
minimizing the differences between the observed and surface corrected model frequencies would show differences at the low frequency end. This arises from how the model frequencies evolve as the model itself evolves. It turns out that if the shape of the assumed surface correction is not correct, then one ends up with either under or over evolved model as best fitting model (depending on whether assumed surface correction over corrects or under corrects).
When only frequency ratios are used as seismic constraints, a much larger difference between the observed and uncorrected theoretical frequencies is obtained at the lower frequency end. This is expected since frequency ratios do not carry information about the surface layers. This has a stronger impact on the radius and density. It is for this reason we added the large separation calculated from $l=0$ mode frequencies.

\section{Summary}
\label{con}

We investigated the internal systematics in mean density, radius,  mass, and age arising from changes in particular physical aspects of stellar models. We did so based on the analysis of stellar model grids constructed as uniformly as possible and only varying the input physics being considered. However, internal systematics arising from the uncertainty in solar metallicity mixture will contain contributions from their respective opacities. Moreover, we also assessed the internal systematics arising from the use of different surface correction methods in forward modelling. 

We found internal systematics from the uncertainty in solar metallicity mixture to be comparable to the statistical uncertainties. Specifically, we found internal systematics of 0.7\%, 0.5\%, 1.4\%, and 6.7\% in mean density, radius, mass, and age, respectively. Relative median statistical uncertainties from using our reference grid (GS98sta) are 0.3\% in density, 0.6\% in radius, 1.6\% in mass, and 7.4\% in age. \citet{Aguirre} found systematic contributions arising the uncertainty in solar metallicity mixture to be 0.3\% in density and radius, 0.6\% in mass and 3.3\% in age. The internal systematics found in this work are approximately twice as large as those found by \citet{Aguirre}. The most probable cause for this difference is the fact that we treat the mixing length parameter, $\alpha_{\rm mlt}$ (see Sect.~\ref{abundance}) as a free parameter.

Concerning the impact of diffusion, we have shown that the inclusion of diffusion in stellar grids of solar-type stars leads to models with significantly lower ages. This is consistent with previous findings \citep{Aguirre,Dotter}. We found internal systematics of 0.5\%, 0.8\%, 2.1\%, and 16\% in mean density, radius, mass, and age, respectively. The internal systematics in age are significantly larger than the corresponding statistical uncertainties.

We assessed the impact of using different surface correction methods on the derived stellar parameters. We found the corrections by Sonoi and BG2 to yield the least internal systematics, namely, 0.9\% and 0.8\%  in radius, 2.0\% and 1.7\% in mass, and 8.2\% and 7.3\% in age, respectively. These internal systematics are comparable to the statistical uncertainties. KJ performs satisfactorily for our sample (see discussion in Sect.~\ref{freq}), while BG1 yields the largest internal systematics as well as the largest biases for stellar radius and mass. We found stellar masses to be overestimated when using the KJ and BG1 corrections. 

Asteroseismology is proving to be particularly significant for the study of solar-type stars, in great part due to the exquisite data that have been made available by NASA's \textit{Kepler} space telescope. The future looks even brighter, with NASA's TESS and ESA's PLATO space missions promising to revolutionise the field and increase the number of stars with detected oscillations by several orders of magnitude. The information contained in stellar oscillations allows the internal stellar structure to be constrained to unprecedented levels, while also allowing fundamental stellar properties (e.g. mass, radius, and age) to be precisely determined. In anticipation of the flood of observations from future space missions, a number of state-of-the-art asteroseismic techniques for the estimation of fundamental stellar properties are currently being developed and tested. Particular attention is being placed on calibrating the determination of age, due to the strong dependence this quantity has on stellar physics. This work therefore provides a valuable contribution to this communal effort by assessing the systematics on the derived stellar properties that arise from specific changes in the model input physics.

\section*{Acknowledgements}
This work was supported by Funda\c{c}\~{a}o para a Ci\^{e}ncia e a Tecnologia (FCT, Portugal) through national funds (UID/FIS/04434/2013) and by FEDER through COMPETE2020 (POCI-01-0145-FEDER-007672).
BN is supported by Funda\c{c}\~{a}o para a Ci\^{e}ncia e a Tecnologia (FCT, Portugal) under the Grant ID: PD/BD/113744/2015 from PHD:SPACE an FCT PhD program.  MSC is supported by FCT through an Investigador contract of reference IF/00894/2012 and POPH/FSE (EC) by FEDER funding through the program COMPETE.
Funding for the Stellar Astrophysics Centre is provided by The Danish National Research Foundation (Grant agreement no.: DNRF106)
The authors thank the MESA user community for the engaging conversations about MESA.
The AIMS project was developed at the University of Birmingham by Daniel R. Reese 
as one of the deliverables for the SPACEINN network. The SPACEINN network was funded by the European Community's Seventh Framework Programme (FP7/2007-2013) under grant agreement no. 312844. We thank the reviewer for the constructive remarks. 



\bibliographystyle{mnras}
\bibliography{mybib} 

\begin{thebibliography}{}
\makeatletter
\relax
\def\mn@urlcharsother{\let\do\@makeother \do\$\do\&\do\#\do\^\do\_\do\%\do\~}
\def\mn@doi{\begingroup\mn@urlcharsother \@ifnextchar [ {\mn@doi@}
  {\mn@doi@[]}}
\def\mn@doi@[#1]#2{\def\@tempa{#1}\ifx\@tempa\@empty \href
  {http://dx.doi.org/#2} {doi:#2}\else \href {http://dx.doi.org/#2} {#1}\fi
  \endgroup}
\def\mn@eprint#1#2{\mn@eprint@#1:#2::\@nil}
\def\mn@eprint@arXiv#1{\href {http://arxiv.org/abs/#1} {{\tt arXiv:#1}}}
\def\mn@eprint@dblp#1{\href {http://dblp.uni-trier.de/rec/bibtex/#1.xml}
  {dblp:#1}}
\def\mn@eprint@#1:#2:#3:#4\@nil{\def\@tempa {#1}\def\@tempb {#2}\def\@tempc
  {#3}\ifx \@tempc \@empty \let \@tempc \@tempb \let \@tempb \@tempa \fi \ifx
  \@tempb \@empty \def\@tempb {arXiv}\fi \@ifundefined
  {mn@eprint@\@tempb}{\@tempb:\@tempc}{\expandafter \expandafter \csname
  mn@eprint@\@tempb\endcsname \expandafter{\@tempc}}}

\bibitem[\protect\citeauthoryear{{Allen}}{{Allen}}{1976}]{Allen}
{Allen} C.~W.,  1976, {Astrophysical Quantities}

\bibitem[\protect\citeauthoryear{{Aller} \& {Chapman}}{{Aller} \&
  {Chapman}}{1960}]{Aller}
{Aller} L.~H.,  {Chapman} S.,  1960, \mn@doi [ApJ] {10.1086/146943}, \href
  {http://adsabs.harvard.edu/abs/1960ApJ...132..461A} {132, 461}

\bibitem[\protect\citeauthoryear{{Angulo} et~al.,}{{Angulo}
  et~al.}{1999}]{Angulo}
{Angulo} C.,  et~al., 1999, \mn@doi [Nuclear Physics A]
  {10.1016/S0375-9474(99)00030-5}, \href
  {http://adsabs.harvard.edu/abs/1999NuPhA.656....3A} {656, 3}

\bibitem[\protect\citeauthoryear{{Asplund}, {Grevesse}, {Sauval}  \&
  {Scott}}{{Asplund} et~al.}{2009}]{Asplund}
{Asplund} M.,  {Grevesse} N.,  {Sauval} A.~J.,   {Scott} P.,  2009, \mn@doi
  [Annual Review of Astron and Astrophys]
  {10.1146/annurev.astro.46.060407.145222}, \href
  {http://adsabs.harvard.edu/abs/2009ARA%26A..47..481A} {47, 481}

\bibitem[\protect\citeauthoryear{{Ball} \& {Gizon}}{{Ball} \&
  {Gizon}}{2014}]{Ball}
{Ball} W.~H.,  {Gizon} L.,  2014, \mn@doi [A \& A]
  {10.1051/0004-6361/201424325}, \href
  {http://adsabs.harvard.edu/abs/2014A%26A...568A.123B} {568, A123}

\bibitem[\protect\citeauthoryear{{Ball} \& {Gizon}}{{Ball} \&
  {Gizon}}{2017}]{WHG}
{Ball} W.~H.,  {Gizon} L.,  2017, \mn@doi [A \& A]
  {10.1051/0004-6361/201630260}, \href
  {http://adsabs.harvard.edu/abs/2017A%26A...600A.128B} {600, A128}

\bibitem[\protect\citeauthoryear{{Balser}}{{Balser}}{2006}]{Balser}
{Balser} D.~S.,  2006, \mn@doi [Astronomical Journal] {10.1086/508515}, \href
  {http://adsabs.harvard.edu/abs/2006AJ....132.2326B} {132, 2326}

\bibitem[\protect\citeauthoryear{{Basu} \& {Antia}}{{Basu} \&
  {Antia}}{2004}]{Basu}
{Basu} S.,  {Antia} H.~M.,  2004, \mn@doi [ApJL] {10.1086/421110}, \href
  {http://adsabs.harvard.edu/abs/2004ApJ...606L..85B} {606, L85}

\bibitem[\protect\citeauthoryear{{Benomar}, {Masuda}, {Shibahashi}  \&
  {Suto}}{{Benomar} et~al.}{2014}]{Beno}
{Benomar} O.,  {Masuda} K.,  {Shibahashi} H.,   {Suto} Y.,  2014, \mn@doi
  [PASJ] {10.1093/pasj/psu069}, \href
  {http://adsabs.harvard.edu/abs/2014PASJ...66...94B} {66, 94}

\bibitem[\protect\citeauthoryear{{B{\"o}hm-Vitense}}{{B{\"o}hm-Vitense}}{1958}]{Vitense}
{B{\"o}hm-Vitense} E.,  1958, Zeit. Astrophys., \href
  {http://adsabs.harvard.edu/abs/1958ZA.....46..108B} {46, 108}

\bibitem[\protect\citeauthoryear{{Borucki} et~al.,}{{Borucki}
  et~al.}{2010}]{Borucki}
{Borucki} W.~J.,  et~al., 2010, \mn@doi [Science] {10.1126/science.1185402},
  \href {http://adsabs.harvard.edu/abs/2010Sci...327..977B} {327, 977}

\bibitem[\protect\citeauthoryear{{Brand{\~a}o} et~al.,}{{Brand{\~a}o}
  et~al.}{2011}]{Brand}
{Brand{\~a}o} I.~M.,  et~al., 2011, \mn@doi [A \& A]
  {10.1051/0004-6361/201015370}, \href
  {http://adsabs.harvard.edu/abs/2011A%26A...527A..37B} {527, A37}

\bibitem[\protect\citeauthoryear{{Brown}, {Gilliland}, {Noyes}  \&
  {Ramsey}}{{Brown} et~al.}{1991}]{Brown}
{Brown} T.~M.,  {Gilliland} R.~L.,  {Noyes} R.~W.,   {Ramsey} L.~W.,  1991,
  \mn@doi [ApJ] {10.1086/169725}, \href
  {http://adsabs.harvard.edu/abs/1991ApJ...368..599B} {368, 599}

\bibitem[\protect\citeauthoryear{{Campante} et~al.,}{{Campante}
  et~al.}{2015}]{Camp}
{Campante} T.~L.,  et~al., 2015, \mn@doi [ApJ] {10.1088/0004-637X/799/2/170},
  \href {http://adsabs.harvard.edu/abs/2015ApJ...799..170C} {799, 170}

\bibitem[\protect\citeauthoryear{{Campante} et~al.,}{{Campante}
  et~al.}{2016a}]{CTL}
{Campante} T.~L.,  et~al., 2016a, \mn@doi [ApJ] {10.3847/0004-637X/819/1/85},
  \href {http://adsabs.harvard.edu/abs/2016ApJ...819...85C} {819, 85}

\bibitem[\protect\citeauthoryear{{Campante} et~al.,}{{Campante}
  et~al.}{2016b}]{Campante}
{Campante} T.~L.,  et~al., 2016b, \mn@doi [ApJ] {10.3847/0004-637X/830/2/138},
  \href {http://adsabs.harvard.edu/abs/2016ApJ...830..138C} {830, 138}

\bibitem[\protect\citeauthoryear{{Campante}, {Santos}  \&
  {Monteiro}}{{Campante} et~al.}{2017}]{SCM}
{Campante} T.~L.,  {Santos} N.~C.,   {Monteiro} M.~J.~P.~F.~G.,  2017,
  preprint, \href {http://adsabs.harvard.edu/abs/2017arXiv170900645C} {}
  (\mn@eprint {arXiv} {1709.00645})

\bibitem[\protect\citeauthoryear{{Casagrande}, {Flynn}, {Portinari}, {Girardi}
  \& {Jimenez}}{{Casagrande} et~al.}{2007}]{Casagrande}
{Casagrande} L.,  {Flynn} C.,  {Portinari} L.,  {Girardi} L.,   {Jimenez} R.,
  2007, \mn@doi [MNRAS] {10.1111/j.1365-2966.2007.12512.x}, \href
  {http://adsabs.harvard.edu/abs/2007MNRAS.382.1516C} {382, 1516}

\bibitem[\protect\citeauthoryear{{Casagrande} et~al.,}{{Casagrande}
  et~al.}{2014}]{Casa}
{Casagrande} L.,  et~al., 2014, \mn@doi [ApJ] {10.1088/0004-637X/787/2/110},
  \href {http://adsabs.harvard.edu/abs/2014ApJ...787..110C} {787, 110}

\bibitem[\protect\citeauthoryear{{Chaboyer}, {Fenton}, {Nelan}, {Patnaude}  \&
  {Simon}}{{Chaboyer} et~al.}{2001}]{Chaboyer}
{Chaboyer} B.,  {Fenton} W.~H.,  {Nelan} J.~E.,  {Patnaude} D.~J.,   {Simon}
  F.~E.,  2001, \mn@doi [ApJ] {10.1086/323872}, \href
  {http://adsabs.harvard.edu/abs/2001ApJ...562..521C} {562, 521}

\bibitem[\protect\citeauthoryear{{Chiosi} \& {Matteucci}}{{Chiosi} \&
  {Matteucci}}{1982}]{Chiosi}
{Chiosi} C.,  {Matteucci} F.~M.,  1982, A \& A, \href
  {http://adsabs.harvard.edu/abs/1982A%26A...105..140C} {105, 140}

\bibitem[\protect\citeauthoryear{{Christensen-Dalsgaard} \&
  {Thompson}}{{Christensen-Dalsgaard} \& {Thompson}}{1997}]{Dalsgaard}
{Christensen-Dalsgaard} J.,  {Thompson} M.~J.,  1997, \mn@doi [MNRAS]
  {10.1093/mnras/284.3.527}, \href
  {http://adsabs.harvard.edu/abs/1997MNRAS.284..527C} {284, 527}

\bibitem[\protect\citeauthoryear{{Christensen-Dalsgaard}, {Dappen}  \&
  {Lebreton}}{{Christensen-Dalsgaard} et~al.}{1988}]{Dalsgaard1}
{Christensen-Dalsgaard} J.,  {Dappen} W.,   {Lebreton} Y.,  1988, \mn@doi
  [Nature] {10.1038/336634a0}, \href
  {http://adsabs.harvard.edu/abs/1988Natur.336..634C} {336, 634}

\bibitem[\protect\citeauthoryear{{Christensen-Dalsgaard}, {Proffitt}  \&
  {Thompson}}{{Christensen-Dalsgaard} et~al.}{1993}]{Christensen}
{Christensen-Dalsgaard} J.,  {Proffitt} C.~R.,   {Thompson} M.~J.,  1993,
  \mn@doi [ApJ] {10.1086/186725}, \href
  {http://adsabs.harvard.edu/abs/1993ApJ...403L..75C} {403, L75}

\bibitem[\protect\citeauthoryear{{Christensen-Dalsgaard}, {Monteiro}, {Rempel}
  \& {Thompson}}{{Christensen-Dalsgaard} et~al.}{2011}]{Thomps}
{Christensen-Dalsgaard} J.,  {Monteiro} M.~J.~P.~F.~G.,  {Rempel} M.,
  {Thompson} M.~J.,  2011, \mn@doi [MNRAS] {10.1111/j.1365-2966.2011.18460.x},
  \href {http://adsabs.harvard.edu/abs/2011MNRAS.414.1158C} {414, 1158}

\bibitem[\protect\citeauthoryear{{Cyburt}, {Fields}  \& {Olive}}{{Cyburt}
  et~al.}{2003}]{Cyburt}
{Cyburt} R.~H.,  {Fields} B.~D.,   {Olive} K.~A.,  2003, \mn@doi [Physics
  Letters B] {10.1016/j.physletb.2003.06.026}, \href
  {http://adsabs.harvard.edu/abs/2003PhLB..567..227C} {567, 227}

\bibitem[\protect\citeauthoryear{{Davies} et~al.,}{{Davies} et~al.}{2016}]{Dav}
{Davies} G.~R.,  et~al., 2016, \mn@doi [MNRAS] {10.1093/mnras/stv2593}, \href
  {http://adsabs.harvard.edu/abs/2016MNRAS.456.2183D} {456, 2183}

\bibitem[\protect\citeauthoryear{{Deheuvels} \& {Michel}}{{Deheuvels} \&
  {Michel}}{2011}]{Deheuvels}
{Deheuvels} S.,  {Michel} E.,  2011, \mn@doi [A \& A]
  {10.1051/0004-6361/201117232}, \href
  {http://adsabs.harvard.edu/abs/2011A%26A...535A..91D} {535, A91}

\bibitem[\protect\citeauthoryear{{Deheuvels} et~al.,}{{Deheuvels}
  et~al.}{2014}]{Deheuvels2014}
{Deheuvels} S.,  et~al., 2014, \mn@doi [A \& A] {10.1051/0004-6361/201322779},
  \href {http://adsabs.harvard.edu/abs/2014A%26A...564A..27D} {564, A27}

\bibitem[\protect\citeauthoryear{{Dotter}, {Conroy}, {Cargile}  \&
  {Asplund}}{{Dotter} et~al.}{2017}]{Dotter}
{Dotter} A.,  {Conroy} C.,  {Cargile} P.,   {Asplund} M.,  2017, \mn@doi [ApJ]
  {10.3847/1538-4357/aa6d10}, \href
  {http://adsabs.harvard.edu/abs/2017ApJ...840...99D} {840, 99}

\bibitem[\protect\citeauthoryear{{Dziembowski}, {Paterno}  \&
  {Ventura}}{{Dziembowski} et~al.}{1988}]{Dziembowski}
{Dziembowski} W.~A.,  {Paterno} L.,   {Ventura} R.,  1988, A \& A, \href
  {http://adsabs.harvard.edu/abs/1988A%26A...200..213D} {200, 213}

\bibitem[\protect\citeauthoryear{{Ferguson}, {Alexander}, {Allard}, {Barman},
  {Bodnarik}, {Hauschildt}, {Heffner-Wong}  \& {Tamanai}}{{Ferguson}
  et~al.}{2005}]{Ferguson}
{Ferguson} J.~W.,  {Alexander} D.~R.,  {Allard} F.,  {Barman} T.,  {Bodnarik}
  J.~G.,  {Hauschildt} P.~H.,  {Heffner-Wong} A.,   {Tamanai} A.,  2005,
  \mn@doi [ApJ] {10.1086/428642}, \href
  {http://adsabs.harvard.edu/abs/2005ApJ...623..585F} {623, 585}

\bibitem[\protect\citeauthoryear{{Figueira}, {Faria}, {Adibekyan}, {Oshagh}  \&
  {Santos}}{{Figueira} et~al.}{2016}]{Figueira}
{Figueira} P.,  {Faria} J.~P.,  {Adibekyan} V.~Z.,  {Oshagh} M.,   {Santos}
  N.~C.,  2016, \mn@doi [Origins of Life and Evolution of the Biosphere]
  {10.1007/s11084-016-9490-5}, \href
  {http://adsabs.harvard.edu/abs/2016OLEB...46..385F} {46, 385}

\bibitem[\protect\citeauthoryear{{Foreman-Mackey}, {Hogg}, {Lang}  \&
  {Goodman}}{{Foreman-Mackey} et~al.}{2013}]{Foreman}
{Foreman-Mackey} D.,  {Hogg} D.~W.,  {Lang} D.,   {Goodman} J.,  2013, \mn@doi
  [PASP] {10.1086/670067}, \href
  {http://adsabs.harvard.edu/abs/2013PASP..125..306F} {125, 306}

\bibitem[\protect\citeauthoryear{{Goldreich}, {Murray}, {Willette}  \&
  {Kumar}}{{Goldreich} et~al.}{1991}]{Goldreich}
{Goldreich} P.,  {Murray} N.,  {Willette} G.,   {Kumar} P.,  1991, \mn@doi
  [ApJ] {10.1086/169858}, \href
  {http://adsabs.harvard.edu/abs/1991ApJ...370..752G} {370, 752}

\bibitem[\protect\citeauthoryear{Gough}{Gough}{1990}]{Gough1990}
Gough D.,  1990, Comments on helioseismic inference, in Progress of Seismology
  of the Sun and Stars.
Springer Berlin Heidelberg, pp Lecture Notes in Physics, Vol 367, 283,
  \mn@doi{10.1007/3-540-53091-6_93}

\bibitem[\protect\citeauthoryear{{Gregory}}{{Gregory}}{2005}]{Gregory}
{Gregory} P.~C.,  2005, {Bayesian Logical Data Analysis for the Physical
  Sciences: A Comparative Approach with `Mathematica' Support}.
Cambridge University Press

\bibitem[\protect\citeauthoryear{{Grevesse} \& {Sauval}}{{Grevesse} \&
  {Sauval}}{1998}]{Grevesse}
{Grevesse} N.,  {Sauval} A.~J.,  1998, \mn@doi [Space Science Reviews]
  {10.1023/A:1005161325181}, \href
  {http://adsabs.harvard.edu/abs/1998SSRv...85..161G} {85, 161}

\bibitem[\protect\citeauthoryear{{Gruberbauer}, {Guenther}, {MacLeod}  \&
  {Kallinger}}{{Gruberbauer} et~al.}{2013}]{Gruberbauer}
{Gruberbauer} M.,  {Guenther} D.~B.,  {MacLeod} K.,   {Kallinger} T.,  2013,
  \mn@doi [MNRAS] {10.1093/mnras/stt1289}, \href
  {http://adsabs.harvard.edu/abs/2013MNRAS.435..242G} {435, 242}

\bibitem[\protect\citeauthoryear{{Guzik} \& {Cox}}{{Guzik} \&
  {Cox}}{1993}]{Guzik}
{Guzik} J.~A.,  {Cox} A.~N.,  1993, \mn@doi [ApJ] {10.1086/172840}, \href
  {http://adsabs.harvard.edu/abs/1993ApJ...411..394G} {411, 394}

\bibitem[\protect\citeauthoryear{{Huber} et~al.,}{{Huber} et~al.}{2013}]{Huber}
{Huber} D.,  et~al., 2013, \mn@doi [ApJ] {10.1088/0004-637X/767/2/127}, \href
  {http://adsabs.harvard.edu/abs/2013ApJ...767..127H} {767, 127}

\bibitem[\protect\citeauthoryear{{Iglesias} \& {Rogers}}{{Iglesias} \&
  {Rogers}}{1996}]{Iglesias}
{Iglesias} C.~A.,  {Rogers} F.~J.,  1996, \mn@doi [ApJ] {10.1086/177381}, \href
  {http://adsabs.harvard.edu/abs/1996ApJ...464..943I} {464, 943}

\bibitem[\protect\citeauthoryear{{Imbriani} et~al.,}{{Imbriani}
  et~al.}{2005}]{Imbriani}
{Imbriani} G.,  et~al., 2005, \mn@doi [EPJ A] {10.1140/epja/i2005-10138-7},
  \href {http://adsabs.harvard.edu/abs/2005EPJA...25..455I} {25, 455}

\bibitem[\protect\citeauthoryear{{Jimenez}, {Flynn}, {MacDonald}  \&
  {Gibson}}{{Jimenez} et~al.}{2003}]{Jimenez}
{Jimenez} R.,  {Flynn} C.,  {MacDonald} J.,   {Gibson} B.~K.,  2003, \mn@doi
  [Science] {10.1126/science.1080866}, \href
  {http://adsabs.harvard.edu/abs/2003Sci...299.1552J} {299, 1552}

\bibitem[\protect\citeauthoryear{{Kjeldsen} \& {Bedding}}{{Kjeldsen} \&
  {Bedding}}{1995}]{Kjeldsen2}
{Kjeldsen} H.,  {Bedding} T.~R.,  1995, \aap, \href
  {http://adsabs.harvard.edu/abs/1995A%26A...293...87K} {293, 87}

\bibitem[\protect\citeauthoryear{{Kjeldsen}, {Bedding}  \&
  {Christensen-Dalsgaard}}{{Kjeldsen} et~al.}{2008}]{Kjeld}
{Kjeldsen} H.,  {Bedding} T.~R.,   {Christensen-Dalsgaard} J.,  2008, \mn@doi
  [ApJL] {10.1086/591667}, \href
  {http://adsabs.harvard.edu/abs/2008ApJ...683L.175K} {683, L175}

\bibitem[\protect\citeauthoryear{{Korn}, {Grundahl}, {Richard}, {Mashonkina},
  {Barklem}, {Collet}, {Gustafsson}  \& {Piskunov}}{{Korn} et~al.}{2007}]{Korn}
{Korn} A.~J.,  {Grundahl} F.,  {Richard} O.,  {Mashonkina} L.,  {Barklem}
  P.~S.,  {Collet} R.,  {Gustafsson} B.,   {Piskunov} N.,  2007, \mn@doi [ApJ]
  {10.1086/523098}, \href {http://adsabs.harvard.edu/abs/2007ApJ...671..402K}
  {671, 402}

\bibitem[\protect\citeauthoryear{{Kunz}, {Fey}, {Jaeger}, {Mayer}, {Hammer},
  {Staudt}, {Harissopulos}  \& {Paradellis}}{{Kunz} et~al.}{2002}]{Kunz}
{Kunz} R.,  {Fey} M.,  {Jaeger} M.,  {Mayer} A.,  {Hammer} J.~W.,  {Staudt} G.,
   {Harissopulos} S.,   {Paradellis} T.,  2002, \mn@doi [ApJ] {10.1086/338384},
  \href {http://adsabs.harvard.edu/abs/2002ApJ...567..643K} {567, 643}

\bibitem[\protect\citeauthoryear{{Lebreton} \& {Goupil}}{{Lebreton} \&
  {Goupil}}{2014}]{Lebreton}
{Lebreton} Y.,  {Goupil} M.~J.,  2014, \mn@doi [A \& A]
  {10.1051/0004-6361/201423797}, \href
  {http://adsabs.harvard.edu/abs/2014A%26A...569A..21L} {569, A21}

\bibitem[\protect\citeauthoryear{{Lodders} \& {Palme}}{{Lodders} \&
  {Palme}}{2009}]{Lodders}
{Lodders} K.,  {Palme} H.,  2009, Meteoritics and Planetary Science Supplement,
  \href {http://adsabs.harvard.edu/abs/2009M%26PSA..72.5154L} {72, 5154}

\bibitem[\protect\citeauthoryear{{Lund} et~al.,}{{Lund} et~al.}{2017}]{Lund}
{Lund} M.~N.,  et~al., 2017, \mn@doi [ApJ] {10.3847/1538-4357/835/2/172}, \href
  {http://adsabs.harvard.edu/abs/2017ApJ...835..172L} {835, 172}

\bibitem[\protect\citeauthoryear{{Maeder} \& {Meynet}}{{Maeder} \&
  {Meynet}}{2000}]{Maeder}
{Maeder} A.,  {Meynet} G.,  2000, \mn@doi [Annual Review of Astron and
  Astrophys] {10.1146/annurev.astro.38.1.143}, \href
  {http://adsabs.harvard.edu/abs/2000ARA%26A..38..143M} {38, 143}

\bibitem[\protect\citeauthoryear{Mamajek}{Mamajek}{2012}]{Mamajek}
Mamajek E.~E.,  2012, ApJL, 754, L20

\bibitem[\protect\citeauthoryear{{Marcy} et~al.,}{{Marcy} et~al.}{2014}]{Marcy}
{Marcy} G.~W.,  et~al., 2014, \mn@doi [ApJS] {10.1088/0067-0049/210/2/20},
  \href {http://adsabs.harvard.edu/abs/2014ApJS..210...20M} {210, 20}

\bibitem[\protect\citeauthoryear{{Mathur} et~al.,}{{Mathur}
  et~al.}{2012}]{Mathur}
{Mathur} S.,  et~al., 2012, \mn@doi [ApJ] {10.1088/0004-637X/749/2/152}, \href
  {http://adsabs.harvard.edu/abs/2012ApJ...749..152M} {749, 152}

\bibitem[\protect\citeauthoryear{{Metcalfe} et~al.,}{{Metcalfe}
  et~al.}{2010}]{Thompson}
{Metcalfe} T.~S.,  et~al., 2010, \mn@doi [ApJ] {10.1088/0004-637X/723/2/1583},
  \href {http://adsabs.harvard.edu/abs/2010ApJ...723.1583M} {723, 1583}

\bibitem[\protect\citeauthoryear{{Metcalfe} et~al.,}{{Metcalfe}
  et~al.}{2012}]{Still}
{Metcalfe} T.~S.,  et~al., 2012, \mn@doi [APJL] {10.1088/2041-8205/748/1/L10},
  \href {http://adsabs.harvard.edu/abs/2012ApJ...748L..10M} {748, L10}

\bibitem[\protect\citeauthoryear{{Metcalfe} et~al.,}{{Metcalfe}
  et~al.}{2014}]{Metcalfe}
{Metcalfe} T.~S.,  et~al., 2014, \mn@doi [ApJS] {10.1088/0067-0049/214/2/27},
  \href {http://adsabs.harvard.edu/abs/2014ApJS..214...27M} {214, 27}

\bibitem[\protect\citeauthoryear{{Miglio} \& {Montalb{\'a}n}}{{Miglio} \&
  {Montalb{\'a}n}}{2005}]{Monta}
{Miglio} A.,  {Montalb{\'a}n} J.,  2005, \mn@doi [A&A]
  {10.1051/0004-6361:20052988}, \href
  {http://adsabs.harvard.edu/abs/2005A%26A...441..615M} {441, 615}

\bibitem[\protect\citeauthoryear{{Monteiro}, {Christensen-Dalsgaard}  \&
  {Thompson}}{{Monteiro} et~al.}{1996}]{Monteiro}
{Monteiro} M.~J.~P.~F.~G.,  {Christensen-Dalsgaard} J.,   {Thompson} M.~J.,
  1996, A \& A, \href {http://adsabs.harvard.edu/abs/1996A%26A...307..624M}
  {307, 624}

\bibitem[\protect\citeauthoryear{{Mosser} et~al.,}{{Mosser} et~al.}{2013}]{mos}
{Mosser} B.,  et~al., 2013, \mn@doi [A&A] {10.1051/0004-6361/201220435}, 550,
  A126

\bibitem[\protect\citeauthoryear{{Nsamba}, {Monteiro}, {Campante}, {Reese},
  {White}, {Garc\'ia Hern\'andez}  \& {Jiang}}{{Nsamba} et~al.}{2017}]{Ben}
{Nsamba} B.,  {Monteiro} M. J. P. F.~G.,  {Campante} T.~L.,  {Reese} D.~R.,
  {White} T.~R.,  {Garc\'ia Hern\'andez} A.,   {Jiang} C.,  2017, \mn@doi [EPJ
  Web Conf.] {10.1051/epjconf/201716005010}, 160, 05010

\bibitem[\protect\citeauthoryear{{Paxton}, {Bildsten}, {Dotter}, {Herwig},
  {Lesaffre}  \& {Timmes}}{{Paxton} et~al.}{2011}]{Pax1}
{Paxton} B.,  {Bildsten} L.,  {Dotter} A.,  {Herwig} F.,  {Lesaffre} P.,
  {Timmes} F.,  2011, \mn@doi [ApJS] {10.1088/0067-0049/192/1/3}, \href
  {http://adsabs.harvard.edu/abs/2011ApJS..192....3P} {192, 3}

\bibitem[\protect\citeauthoryear{{Paxton} et~al.,}{{Paxton}
  et~al.}{2013}]{Pax2}
{Paxton} B.,  et~al., 2013, \mn@doi [ApJS] {10.1088/0067-0049/208/1/4}, \href
  {http://adsabs.harvard.edu/abs/2013ApJS..208....4P} {208, 4}

\bibitem[\protect\citeauthoryear{{Paxton} et~al.,}{{Paxton}
  et~al.}{2015}]{Pax3}
{Paxton} B.,  et~al., 2015, \mn@doi [ApJS] {10.1088/0067-0049/220/1/15}, \href
  {http://adsabs.harvard.edu/abs/2015ApJS..220...15P} {220, 15}

\bibitem[\protect\citeauthoryear{{Perryman}}{{Perryman}}{2014}]{Perryman2014}
{Perryman} M.,  2014, {The Exoplanet Handbook, Cambridge, UK: Cambridge
  University Press}

\bibitem[\protect\citeauthoryear{{Piau}, {Kervella}, {Dib}  \&
  {Hauschildt}}{{Piau} et~al.}{2011}]{Piau}
{Piau} L.,  {Kervella} P.,  {Dib} S.,   {Hauschildt} P.,  2011, \mn@doi [A \&
  A] {10.1051/0004-6361/201014442}, \href
  {http://adsabs.harvard.edu/abs/2011A%26A...526A.100P} {526, A100}

\bibitem[\protect\citeauthoryear{{Pinsonneault}, {An}, {Molenda-{\.Z}akowicz},
  {Chaplin}, {Metcalfe}  \& {Bruntt}}{{Pinsonneault}
  et~al.}{2012}]{Pinsonneault}
{Pinsonneault} M.~H.,  {An} D.,  {Molenda-{\.Z}akowicz} J.,  {Chaplin} W.~J.,
  {Metcalfe} T.~S.,   {Bruntt} H.,  2012, \mn@doi [ApJS]
  {10.1088/0067-0049/199/2/30}, \href
  {http://adsabs.harvard.edu/abs/2012ApJS..199...30P} {199, 30}

\bibitem[\protect\citeauthoryear{{Pinsonneault} et~al.,}{{Pinsonneault}
  et~al.}{2014}]{Pinso}
{Pinsonneault} M.~H.,  et~al., 2014, \mn@doi [ApJS]
  {10.1088/0067-0049/215/2/19}, \href
  {http://adsabs.harvard.edu/abs/2014ApJS..215...19P} {215, 19}

\bibitem[\protect\citeauthoryear{{Ram{\'{\i}}rez}, {Mel{\'e}ndez}  \&
  {Asplund}}{{Ram{\'{\i}}rez} et~al.}{2009}]{Ram}
{Ram{\'{\i}}rez} I.,  {Mel{\'e}ndez} J.,   {Asplund} M.,  2009, \mn@doi [A \&
  A] {10.1051/0004-6361/200913038}, \href
  {http://adsabs.harvard.edu/abs/2009A%26A...508L..17R} {508, L17}

\bibitem[\protect\citeauthoryear{{Rauer} et~al.,}{{Rauer}
  et~al.}{2014}]{Rauer1}
{Rauer} H.,  et~al., 2014, \mn@doi [Experimental Astronomy]
  {10.1007/s10686-014-9383-4}, \href
  {http://adsabs.harvard.edu/abs/2014ExA....38..249R} {38, 249}

\bibitem[\protect\citeauthoryear{{Rogers} \& {Nayfonov}}{{Rogers} \&
  {Nayfonov}}{2002}]{Rogers}
{Rogers} F.~J.,  {Nayfonov} A.,  2002, \mn@doi [ApJ] {10.1086/341894}, \href
  {http://adsabs.harvard.edu/abs/2002ApJ...576.1064R} {576, 1064}

\bibitem[\protect\citeauthoryear{{Roxburgh} \& {Vorontsov}}{{Roxburgh} \&
  {Vorontsov}}{2003}]{Rox}
{Roxburgh} I.~W.,  {Vorontsov} S.~V.,  2003, \mn@doi [A \& A]
  {10.1051/0004-6361:20031318}, \href
  {http://adsabs.harvard.edu/abs/2003A%26A...411..215R} {411, 215}

\bibitem[\protect\citeauthoryear{{Serenelli} \& {Basu}}{{Serenelli} \&
  {Basu}}{2010}]{Serenelli}
{Serenelli} A.~M.,  {Basu} S.,  2010, \mn@doi [ApJ]
  {10.1088/0004-637X/719/1/865}, \href
  {http://adsabs.harvard.edu/abs/2010ApJ...719..865S} {719, 865}

\bibitem[\protect\citeauthoryear{{Silva Aguirre}, {Ballot}, {Serenelli}  \&
  {Weiss}}{{Silva Aguirre} et~al.}{2011}]{Silva}
{Silva Aguirre} V.,  {Ballot} J.,  {Serenelli} A.~M.,   {Weiss} A.,  2011,
  \mn@doi [A \& A] {10.1051/0004-6361/201015847}, \href
  {http://adsabs.harvard.edu/abs/2011A%26A...529A..63S} {529, A63}

\bibitem[\protect\citeauthoryear{{Silva Aguirre} et~al.,}{{Silva Aguirre}
  et~al.}{2013}]{Kawaler}
{Silva Aguirre} V.,  et~al., 2013, \mn@doi [ApJ] {10.1088/0004-637X/769/2/141},
  \href {http://adsabs.harvard.edu/abs/2013ApJ...769..141S} {769, 141}

\bibitem[\protect\citeauthoryear{{Silva Aguirre} et~al.,}{{Silva Aguirre}
  et~al.}{2015}]{Aguirre}
{Silva Aguirre} V.,  et~al., 2015, \mn@doi [\mnras] {10.1093/mnras/stv1388},
  \href {http://adsabs.harvard.edu/abs/2015MNRAS.452.2127S} {452, 2127}

\bibitem[\protect\citeauthoryear{{Silva Aguirre} et~al.,}{{Silva Aguirre}
  et~al.}{2017}]{Aguirre1}
{Silva Aguirre} V.,  et~al., 2017, \mn@doi [\apj]
  {10.3847/1538-4357/835/2/173}, \href
  {http://adsabs.harvard.edu/abs/2017ApJ...835..173S} {835, 173}

\bibitem[\protect\citeauthoryear{{Sonoi}, {Samadi}, {Belkacem}, {Ludwig},
  {Caffau}  \& {Mosser}}{{Sonoi} et~al.}{2015}]{Sonoi}
{Sonoi} T.,  {Samadi} R.,  {Belkacem} K.,  {Ludwig} H.-G.,  {Caffau} E.,
  {Mosser} B.,  2015, \mn@doi [A \& A] {10.1051/0004-6361/201526838}, \href
  {http://adsabs.harvard.edu/abs/2015A%26A...583A.112S} {583, A112}

\bibitem[\protect\citeauthoryear{{Thoul}, {Bahcall}  \& {Loeb}}{{Thoul}
  et~al.}{1994}]{Thoul}
{Thoul} A.~A.,  {Bahcall} J.~N.,   {Loeb} A.,  1994, \mn@doi [ApJ]
  {10.1086/173695}, \href {http://adsabs.harvard.edu/abs/1994ApJ...421..828T}
  {421, 828}

\bibitem[\protect\citeauthoryear{{Townsend} \& {Teitler}}{{Townsend} \&
  {Teitler}}{2013}]{Townsend}
{Townsend} R.~H.~D.,  {Teitler} S.~A.,  2013, \mn@doi [MNRAS]
  {10.1093/mnras/stt1533}, \href
  {http://adsabs.harvard.edu/abs/2013MNRAS.435.3406T} {435, 3406}

\bibitem[\protect\citeauthoryear{{Trampedach} \& {Stein}}{{Trampedach} \&
  {Stein}}{2011}]{Trampedach}
{Trampedach} R.,  {Stein} R.~F.,  2011, \mn@doi [ApJ]
  {10.1088/0004-637X/731/2/78}, \href
  {http://adsabs.harvard.edu/abs/2011ApJ...731...78T} {731, 78}

\bibitem[\protect\citeauthoryear{{Trampedach}, {Stein},
  {Christensen-Dalsgaard}, {Nordlund}  \& {Asplund}}{{Trampedach}
  et~al.}{2014}]{Nordlund}
{Trampedach} R.,  {Stein} R.~F.,  {Christensen-Dalsgaard} J.,  {Nordlund}
  {\AA}.,   {Asplund} M.,  2014, \mn@doi [MNRAS] {10.1093/mnras/stu2084}, \href
  {http://adsabs.harvard.edu/abs/2014MNRAS.445.4366T} {445, 4366}

\bibitem[\protect\citeauthoryear{{Turcotte}, {Richer}, {Michaud}, {Iglesias}
  \& {Rogers}}{{Turcotte} et~al.}{1998}]{Turcotte}
{Turcotte} S.,  {Richer} J.,  {Michaud} G.,  {Iglesias} C.~A.,   {Rogers}
  F.~J.,  1998, \mn@doi [ApJ] {10.1086/306055}, \href
  {http://adsabs.harvard.edu/abs/1998ApJ...504..539T} {504, 539}

\bibitem[\protect\citeauthoryear{{Van Eylen}, {Kjeldsen},
  {Christensen-Dalsgaard}  \& {Aerts}}{{Van Eylen} et~al.}{2012}]{Eylen}
{Van Eylen} V.,  {Kjeldsen} H.,  {Christensen-Dalsgaard} J.,   {Aerts} C.,
  2012, \mn@doi [Astronomische Nachrichten] {10.1002/asna.201211832}, \href
  {http://adsabs.harvard.edu/abs/2012AN....333.1088V} {333, 1088}

\bibitem[\protect\citeauthoryear{{VandenBerg}, {Richard}, {Michaud}  \&
  {Richer}}{{VandenBerg} et~al.}{2002}]{VandenBerg}
{VandenBerg} D.~A.,  {Richard} O.,  {Michaud} G.,   {Richer} J.,  2002, \mn@doi
  [ApJ] {10.1086/339895}, \href
  {http://adsabs.harvard.edu/abs/2002ApJ...571..487V} {571, 487}

\bibitem[\protect\citeauthoryear{{White} et~al.,}{{White} et~al.}{2017}]{white}
{White} T.~R.,  et~al., 2017, \mn@doi [\aap] {10.1051/0004-6361/201628706},
  \href {http://adsabs.harvard.edu/abs/2017A%26A...601A..82W} {601, A82}

\makeatother
\end{thebibliography}







\bsp	
\label{lastpage}
\end{document}